\documentclass[prb,superscriptaddress,floatfix,twocolumn,showpacs,amsfonts,amsmath,amssymb,final]{revtex4-1}

\usepackage{graphicx} % include figure files

\usepackage{dcolumn} % align table columns on decimal point

\usepackage{bm} % bold math

\begin{document}

\title{Three dimensional generalization of the $J_1$-$J_2$ Heisenberg model on a square lattice and role of the interlayer coupling $J_c$}

\author{Michael Holt}

\author{Oleg P. Sushkov}

\affiliation{School of Physics, University of New South Wales, Kensington 2052, Sydney NSW, Australia}

\author{Daniel Stanek}

\affiliation{Lehrstuhl f\"{u}r Theoretische Physik I, Technische  Universit\"{a}t Dortmund, Otto-Hahn Stra\ss{}e 4, 44221 Dortmund, Germany}

\author{G\"otz S. Uhrig}

\affiliation{Lehrstuhl f\"{u}r Theoretische Physik I, Technische  Universit\"{a}t Dortmund, Otto-Hahn Stra\ss{}e 4, 44221 Dortmund, Germany}

\date{\rm\today}

\begin{abstract}

{A possibility to describe magnetism in the iron pnictide parent compounds in terms of the  two-dimensional frustrated Heisenberg $J_1$-$J_2$ model has been actively  discussed recently. However, recent neutron scattering data has shown that the pnictides have a relatively  large spin  wave  dispersion in the direction perpendicular to the planes. This indicates that the third dimension is very important. Motivated by this observation  we study the $J_1$-$J_2$-$J_c$ model that is the three dimensional generalization of the  $J_1$-$J_2$ Heisenberg model for $S = 1/2$ and $S = 1$. Using self-consistent spin  wave theory we present a detailed description of the staggered magnetization and magnetic excitations in the collinear state. We find that the introduction of the interlayer coupling $J_c$ suppresses the quantum fluctuations and strengthens the long range ordering. In the $J_1$-$J_2$-$J_c$ model, we find two qualitatively distinct scenarios for how the collinear phase becomes unstable upon increasing $J_1$. Either the magnetization or one of the spin wave velocities vanishes. For $S = 1/2$ renormalization due to quantum fluctuations is significantly stronger than for $S=1$, in particular close to the quantum phase transition. Our findings for the $J_1$-$J_2$-$J_c$ model are of general theoretical interest, however, the results show that it is unlikely that the model is relevant to undoped pnictides.}

\end{abstract}

\maketitle

\section{Introduction}

\label{sec:introduction}

Over the last two decades, there has been considerable interest in the two dimensional (2D) $S = 1/2$ Heisenberg antiferromagnet with frustrating interactions. One of the most widely studied models is the square lattice $J_1$-$J_2$ model, with both nearest neighbour $J_1$ and second nearest neighbour $J_2$ antiferromagnetic interactions. The Hamiltonian is

\begin{equation}
\label{eq:hamilton}
H=J_1\sum_{\langle i,j\rangle} {\bf S}_i\cdot {\bf S}_j + J_2\sum_{\langle\langle i,j\rangle\rangle} {\bf S}_i\cdot {\bf S}_j
\end{equation}
where $\langle i,j\rangle$ stands for summation over nearest neighbors (NN) while $\langle\langle i,j\rangle\rangle$ stands for summation over next-nearest neighbors (NNN). Betts and Oitmaa \cite{betts77} were the first to point out that there is a finite long range order in the two dimensional Heisenberg model at zero temperature. Early studies \cite{manou91} showed that the ground state of the pure $J_1$ model has N\'eel order reduced by quantum fluctuations. The N\'{e}el order is destabilized with increasing $J_2$ and at some critical value of $J_2/J_1$ a phase transition to a quantum disordered phase occurs. On the other hand, for large $J_2/J_1$ the system will order in a “stripe-like” fashion of alternating rows (or columns) of spins up and down. The long-range magnetic order is reduced by quantum fluctuations. As $J_2/J_1$ is reduced the collinear phase will become unstable at some critical ratio. There is substantial evidence; see Refs.\ \onlinecite{singh99b, kotov99b, sushk01} and references within, that the ground state of the quantum disordered phase has no long-range magnetic order and is dominated by short-range singlet (dimer) formation for $0.4$ $<$ $J_2$/$J_1$ $<$ $0.6$ for $S=1/2$. The stability of such a configuration implies that the lattice symmetry is spontaneously broken and the ground state is fourfold degenerate. 

Using series expansion and mean-field spin wave theory methods, Singh et al.\ studied the excitation spectra of the square lattice $J_1$-$J_2$ Heisenberg antiferromagnet \cite{singh03}. They showed the excitation spectra is gapless at only two symmetry related points of the Brillouin zone $(0,0)$ and $(0,\pi)$, whereas the accidental degeneracies at $(\pi,\pi)$ and $(\pi,0)$ are lifted by the `order by disorder' effect \cite{shend82}, where the quantum fluctuations select a collinear ground state. Furthermore, they found the ratio of the spin-wave velocities along the $x$ and $y$ directions depends sharply on the $J_2/J_1$ ratio. 

Besides being of general theoretical interest the $J_1$-$J_2$ model is relevant to the real layered magnetic materials \cite{melzi00, rosner02}. However, the real materials are not strictly two dimensional and contain a small interlayer coupling $J_{c}$. For example, Rosner et. al. \cite{rosner02} found that $J_{c}/J_{1} \approx 0.07$ for Li$_{2}$VOSiO$_{4}$, which can be described by a square lattice $J_1$-$J_2$ model with large $J_{2}$ \cite{melzi00, rosner02}. This provides motivation for studies of a three dimensional extension of the $J_1$-$J_2$ model. Such an extension for $S=1/2$  has been recently studied using coupled-cluster and rotation-invariant Green's function methods \cite{schma06}, a version of effective field theory \cite{nunes10} as well as different kinds of spin wave approaches \cite{nunes10a, maj10}. In particular these studies focused on the influence of $J_{c}$ on the existence of an intermediate quantum disordered phase at $J_2/J_1 \approx 0.5$. Schmalfu\ss{} et.al. \cite{schma06} found that upon increasing the interlayer coupling $J_{c} > 0$ the intermediate phase disappears at $J_{c} \approx (0.4 - 0.6)J_{2}$. Our interest in the 3D model is mainly motivated by the discovery of superconductivity in the iron pnictides \cite{kamih08}.

\begin{figure}

\begin{center}

\includegraphics[trim  = 0mm 0mm 0mm 0mm, width=0.8\columnwidth,clip]{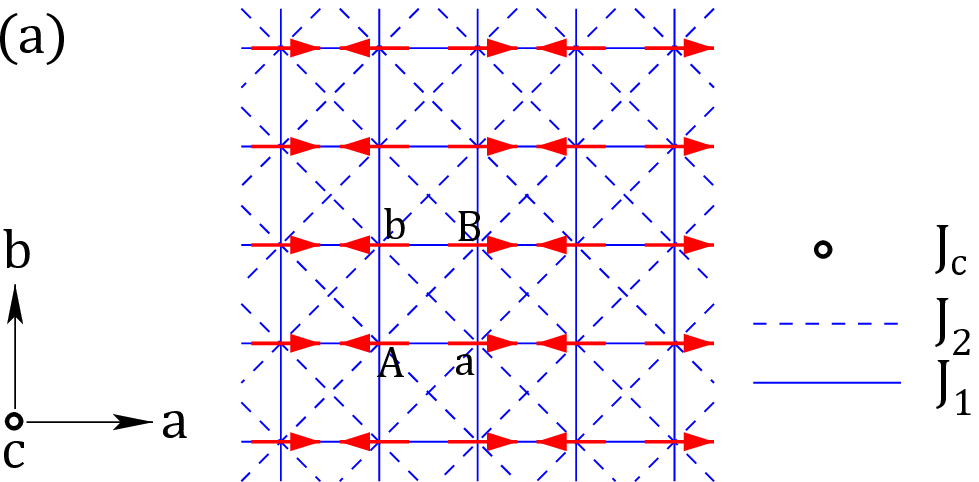}

\includegraphics[trim  = 0mm 0mm 0mm 0mm, width=0.8\columnwidth,clip]{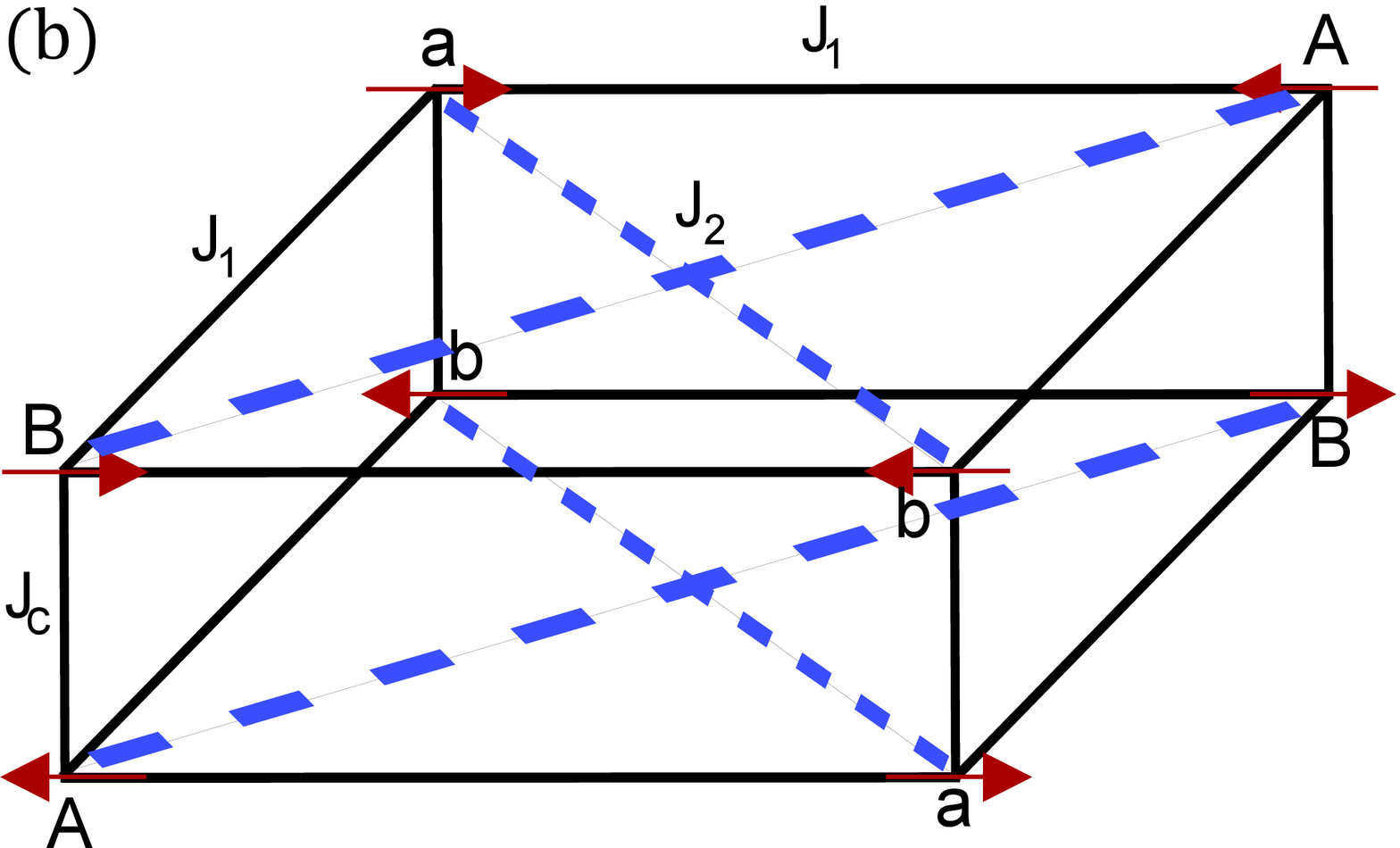}

\end{center}

\caption{(color online) (a) Schematic diagram of the three dimensional spin ordering in the Fe pnictides. Here we show the $a-b$ plane for the considered model with nearest neighbour coupling $J_1$ and next nearest neighbour coupling $J_2$. The interplane coupling $J_c$ is directed into the page. In addition we show in (b) the real space positions of the lattice points $A$, $a$, $B$, and $b$ which make up the unit cell.}

\label{fig:magnet}

\end{figure}

Parent pnictides demonstrate alternating spin stripes and therefore it is quite natural to assume that the collinear phase of $J_1$-$J_2$ model describes the system. Since their discovery, many investigations have been focused on understanding the magnetic properties of the pnictide parent compounds \cite{fang08, xu08, ma09, uhrig09a, bao09, pulik10, si08, wu08}. Magnetic long range order has been established in LaOFeAs and Sr(Ba,Ca)Fe$_2$As$_2$ using neutron scattering \cite{zhao08a, zhao09, ewing08, mcque08, cruz08}, muon spin resonance ($\mu$SR) \cite{drew08} and M\"ossbauer spectroscopy \cite{klaus08, luetk09}. The neutron studies reveal the parent compounds display a columnar antiferromagnetic ordering with a staggered magnetic moment of $(0.3-0.4)\mu_B$ in LaOFeAs and $(0.8-1.01)\mu_B$ in Sr(Ba,Ca)Fe$_2$As$_2$. In this columnar arrangement, stripes of parallel spin order along the $b$ axis and antiferromagnetically along the $a$ and $c$ axes \cite{cruz08,huang08}; see Fig.\ \ref{fig:magnet}.

The spin wave velocities along the $a$ and $c$ axes have been measured in neutron scattering studies of SrFe$_2$As$_2$ \cite{zhao08a} and BaFe$_2$As$_2$ \cite{ewing08} with $v_a \approx$ 205 meV and $v_c \approx$ 45 meV. Upon lowering the temperature the parent pnictide compounds undergo a structural lattice distortion from a tetragonal to an orthorhombic structure. The orthorhombic distortion is very small, and is of order a fraction of one per cent \cite{zhao08b}. The structural transition happens at a temperature slightly higher or equal to the magnetic ordering temperature. This coincidence naturally suggests that the structural transition can be driven by the spontaneous violation of the $Z_2$ symmetry (nematic transition) in the spin stripe phase \cite{xu08,fang08}. Note that this suggestion is based purely on symmetry arguments and is valid for both the ``localized'' and ``itinerant'' paradigm. Let us call this scenario, scenario A. The structural transition can also be fully independent of the spin structure (scenario B). Moreover, the transition can drive the spin structure (scenario C). In the present work we ignore the very small orthorhombic distortion and consider the "isotropic" model i.e. a tetragonal lattice. In relation to the iron pnictides this approach makes sense in cases A and B, and it is not justified in case C.

In the tetragonal phase, the Fe sites form square planar arrays, such that the sites of adjacent planes lie above each other; see  Fig.\ \ref{fig:magnet}. Recently, there has been debate concerning the spin wave velocity along the $b$-axis i.e. along the spin stripes \cite{zhao08a, ong09}. On the one hand direct neutron scattering data from twinned samples \cite{zhao08a} indicate the value of $v_b$ is comparable with $v_a$. On the other hand analysis \cite{ong09} of the NMR relaxation rate indicates an order of magnitude smaller value of $v_b$, $v_b \sim$ 10-30 meV. The small value of $v_b$ implies that the system is close to a quantum critical point. Therefore from now on we will refer to the small $v_b$ scenario as the \emph{critical} one and the large $v_b$ scenario as the \emph{non-critical} scanario.

Band structure calculations \cite{dong08b, yin08} have shown that columnar antiferromagnetic ordering is the most stable structure. On the one hand, band structure results  indicate a local moment of up to $2.3\mu_B$ per Fe site \cite{ma09, yin08, cao08}. This value is too large compared to experiment. This has led to the suggestion that the ordered moment might be strongly renormalized by magnetic fluctuations, as described by the frustrated 2D $J_1$-$J_2$ model \cite{uhrig09a, si08, ong09, yao08}. This suggestion implies the critical scenario because of very strong quantum fluctuations. On the other hand, there are studies suggesting that the smallness of the staggered magnetic moment can be explained by electronic effects such as hybridization and spin-orbit coupling \cite{wu08, yildr08}.

Motivated by what is currently known about the pnictides the goal of the present work is three-fold. First, we provide a quantitative theory for the magnetic excitations based on a minimal spin model, namely the $J_1$-$J_2$ Heisenberg model with interlayer coupling $J_c$ for $S = 1/2$ and $S=1$. In particular, we will show the interlayer exchange coupling suppresses the quantum fluctuations, strengthens the staggered moment, and hence dramatically increases the stability of the columnar phase. Furthermore, it has been recently suggested that the strong reduction of the magnetic moment as seen in 2D is not possible in 3D \cite{smera10} because the 3D coupling cuts the logarithmic divergence of quantum flucuations. In this paper we will show that a considerable reduction of the staggered moment is still possible for $S=1/2$ for small values of $J_c$ while a significant renormalization for $S=1$ is only possible for extreme fine-tuning.

Second, we discuss the phase diagram and the existence of a dimerized phase for the $J_1$-$J_2$-$J_c$ Heisenberg model. Indeed we find a dependence of the point where the staggered magnetization vanishes as a function of interlayer coupling $J_c$. In particular, two different  instabilities of the columnar phase appear: Either the long-range order  or one of the spin wave velocities vanishes. Note that in three dimensions the vanishing of one of the spin wave velocities does \emph{not} imply the vanishing of the staggered magnetization. We will discuss these two kinds of instabilities in a forthcoming publication in more detail \cite{stane10}.

Third, we make quantitative predictions of the dispersion and spin wave velocities parallel ($v_a$), perpendicular ($v_b$), and through ($v_c$) the magnetic stripes; the ratio $v_b/v_a$ is a useful way to determine the degree of magnetic frustration $J_1$/$J_2$. The spin wave spectra over the full Brillouin zone show dramatic differences between a system deep in the columnar phase and one close to a quantum critical point. This provides a robust experimental way to distinguish the critical and non-critical scenarios presented above. In addition, we will compare our model with experimental evidence and discuss whether the $J_1$-$J_2$-$J_c$ is an appropriate model for describing magnetism in the pnictide parent compounds.

\section{Model and Methods}

\label{sec:model}

It is well known that the parent iron pnictides are not simple Mott-Hubbard insulators, but rather are bad metals with a very small Fermi surface. On the one hand this supports an itinerant picture for the compounds while on the other hand there are many experimental indications for localized moments. On a pure theoretical level it is clear that the Hubbard on-site repulsion is rather large and therefore even an itinerant system must be close to the Mott-Hubbard regime. Most likely the truth is somewhere in the middle since there are localized and delocalized degrees of freedom, and it is not clear yet how to combine these two descriptions.

In the present work we disregard the itinerant degrees of freedom and consider a model of well-localized spins described by a Heisenberg Hamiltonian consisting of effective in-plane nearest neighbour $J_1$, next  nearest neighbour $J_2$ and interlayer $J_c$ exchange interactions
\begin{equation}
\label{eq:3dhamilton}
H=J_1\sum_{\langle i,j\rangle} {\bf S}_i\cdot {\bf S}_j + J_2\sum_{\langle\langle i,j\rangle\rangle} {\bf S}_i\cdot {\bf S}_j + J_c\sum_{| i, j|} {\bf S}_i\cdot {\bf S}_j
\end{equation}
where $\langle i,j\rangle$ and $\langle\langle i,j\rangle\rangle$ correspond to summation over nearest-neighbour (NN) and next-nearest neighbour (NNN) pairs in the plane and $| i, j|$ corresponds to nearest neighbour pairs between the planes, see Fig.~\ref{fig:magnet}. 

Since we want to describe the observed spin-stripe phase in the parent pnictides we are interested in the region of parameters of the effective Heisenberg model that supports the phase. It is well known that this phase is stable at sufficiently large $J_{2}$, $J_{2} > 0.5 J_1$. The particular values of $J_1$, $J_2$, $J_c$ can be determined from comparison with known magnetic excitation spectra and we will discuss this issue later in the paper. Note that we consider the ``isotropic'' Heisenberg model, the value of $J_1$ is the same in both directions, along the spin stripes and perpendicular to the stripes so that we consider the entire stripe ordering to be purely spontaneous.

\begin{figure}

    \begin{center}

     \includegraphics[width=0.875\columnwidth,clip]{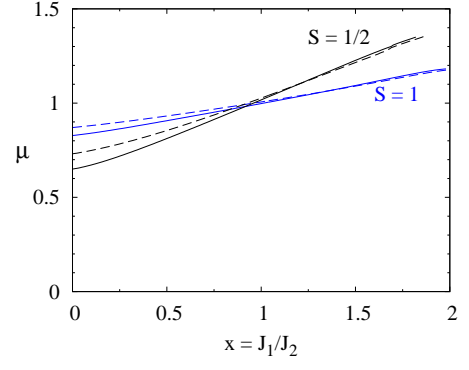}

     \includegraphics[width=0.875\columnwidth,clip]{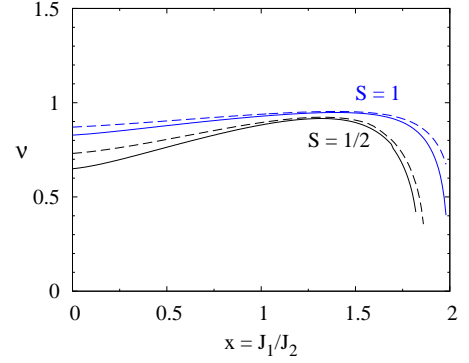}

     \includegraphics[width=0.875\columnwidth,clip]{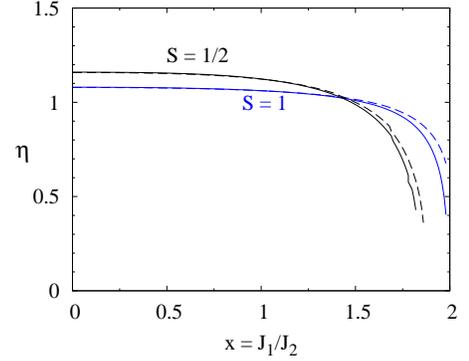}

     \includegraphics[width=0.875\columnwidth,clip]{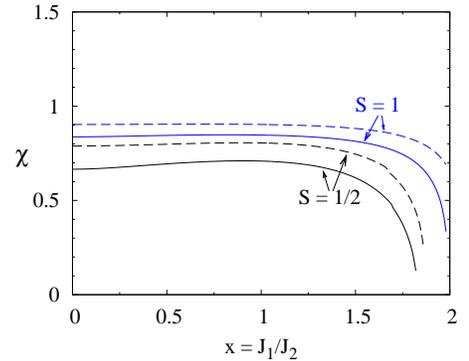}

    \end{center}

    \caption{(color online) Quantum correction parameters $\mu$, $\nu$, $\eta$, $\chi$ as a function of $J_1/J_2$. Calculations were performed by numerical iteration for a range of $y$ values for both $S=1/2$ and $S=1$. Here we show the results for $y = 0.01$ (solid) and $0.10$ (dashed).}

    \label{fig:quantum}

\end{figure}

We calculate the sublattice magnetization and magnetic excitations using self-consistent spin wave theory for $S = 1/2$ and $S = 1$. This method has already been shown to work well for $S = 1/2$ and $S = 1$ in the columnar phase for the two dimensional case \cite{singh03, uhrig09a, oitma96c}. We have used the Dyson-Maleev \cite{dyson56a, malee57} as well as the Schwinger boson representation \cite{auerb94} which on the level of self-consistent mean field theory yield the same results for $T = 0$. The boson operators $A_i$, $a_i$, $B_j$, $b_j$ on the respective sublattices, see Fig.~\ref{fig:magnet}, are introduced in the usual way by performing a Dyson-Maleev \cite{dyson56a} transformation of the spin operators.
\begin{eqnarray}
\label{eq:dyson}
S_i^z &=& S - {a_i}^\dagger a_i\nonumber\\
S_i^\dagger &=& (2S - {a_i}^\dagger {a_i}){a_i} \nonumber\\
S_i^- &=& {a_i}^\dagger \nonumber\\
S_j^z &=& -S + {b_j}^\dagger b_j\nonumber\\
S_j^\dagger &=&-{b_j}^\dagger (2S - {b_j}^\dagger {b_j})\nonumber \\
S_j^- &=& -{b_j} 
\end{eqnarray}
The definition of $A$ and $B$ is similar to $a$ and $b$. Using \eqref{eq:dyson} the Hamiltonian \eqref{eq:3dhamilton} may be presented to quartic order in the operators $A$, $a$, $B$ and $b$
\begin{widetext}
\begin{eqnarray}
\label{eq:quarthamilton}
H &=& H^{Aa} + H^{Bb} + H^{Ab} + H^{aB} + H^{ab} + H^{AB}\\
H^{Aa} &=& - J_1S^2 + J_1\sum_{\langle i,j\rangle}\left\{S({A_i}^\dagger A_i + {a_j}^\dagger a_j - A_i a_j - {A_i}^\dagger {a_j}^\dagger) + \frac{1}{2}({A_i}^\dagger A_i A_i a_j - 2{A_i}^\dagger A_i {a_j}^\dagger a_j + {A_i}^\dagger {a_j}^\dagger {a_j}^\dagger {a_j})\right\}\nonumber \\
H^{Bb} &=& - J_1S^2 + J_1\sum_{\langle i,j\rangle}\left\{S({B_i}^\dagger B_i + {b_j}^\dagger b_j - B_i b_j - {B_i}^\dagger {b_j}^\dagger) + \frac{1}{2}({B_i}^\dagger B_i B_i b_j - 2{B_i}^\dagger B_i {b_j}^\dagger b_j + {B_i}^\dagger {b_j}^\dagger {b_j}^\dagger {b_j})\right\}\nonumber\\
H^{Ab} &=& J_1S^2 - J_1\sum_{\langle i,j\rangle}\left\{S({A_i}^\dagger A_i + {a_j}^\dagger a_j - A_i {b_j}^\dagger - {A_i}^\dagger {b_j}) +\frac{1}{2}({A_i}^\dagger A_i A_i b_j - 2{A_i}^\dagger A_i {b_j}^\dagger b_j + {A_i}^\dagger {b_j}^\dagger {b_j}^\dagger {b_j})\right\}\nonumber\\
H^{aB} &=& J_1S^2 - J_1\sum_{\langle i,j\rangle}\left\{S({a_i}^\dagger a_i + {B_j}^\dagger B_j - a_i {B_j}^\dagger - {a_i}^\dagger {B_j}) +\frac{1}{2}({a_i}^\dagger a_i a_i B_j - 2{a_i}^\dagger a_i {B_j}^\dagger B_j + {a_i}^\dagger {B_j}^\dagger {B_j}^\dagger {B_j})\right\}\nonumber\\
H^{ab} &=& -J_2S^2 + J_2\sum_{\langle\langle i,j\rangle\rangle}\left\{S({a_i}^\dagger a_i + {b_j}^\dagger b_j - a_i {b_j} - {a_i}^\dagger {b_j}^\dagger)
+\frac{1}{2}({a_i}^\dagger a_i a_i b_j - 2{a_i}^\dagger a_i {b_j}^\dagger b_j + {a_i}^\dagger {b_j}^\dagger {b_j}^\dagger {b_j})\right\}\nonumber\\
&-&J_cS^2 + J_c\sum_{| i, j|}\left\{S({a_i}^\dagger a_i + {b_j}^\dagger b_j - a_i {b_j} - {a_i}^\dagger {b_j}^\dagger)+ \frac{1}{2}({a_i}^\dagger a_i a_i b_j - 2{a_i}^\dagger a_i {b_j}^\dagger b_j + {a_i}^\dagger {b_j}^\dagger {b_j}^\dagger {b_j})\right\}\nonumber\\
H^{AB} &=& -J_2S^2 + J_2\sum_{\langle\langle i,j\rangle\rangle}\left\{S({A_i}^\dagger A_i + {B_j}^\dagger B_j - A_i {B_j} - {A_i}^\dagger {B_j}^\dagger)
+ \frac{1}{2}({A_i}^\dagger A_i A_i B_j - 2{A_i}^\dagger A_i {B_j}^\dagger B_j + {A_i}^\dagger {B_j}^\dagger {B_j}^\dagger {B_j})\right\}\nonumber\\
&-& J_cS^2 + J_c\sum_{| i, j|}\left\{S({A_i}^\dagger A_i + {B_j}^\dagger B_j - A_i {B_j} - {A_i}^\dagger {B_j}^\dagger)+ \frac{1}{2}({A_i}^\dagger A_i A_i B_j - 2{A_i}^\dagger A_i {B_j}^\dagger B_j + {A_i}^\dagger {B_j}^\dagger {B_j}^\dagger {B_j})\right\} \ .\nonumber
\end{eqnarray}
\end{widetext}
We perform the Hartree-Fock mean field decoupling of quartic terms using the following notations.
\begin{eqnarray}
\label{eq:expectation}
f - \frac{1}{2} &=& \langle {A_i}^\dagger A_i\rangle =  \langle {a_i}^\dagger a_i\rangle = \langle {B_j}^\dagger B_j\rangle = \langle {b_j}^\dagger b_j\rangle\nonumber\\
F  &=& \langle {A_i} b_j\rangle = \langle {A_i}^\dagger {b_j}^\dagger\rangle =\langle {a_i} B_j\rangle = \langle {a_i}^\dagger {B_j}^\dagger\rangle\nonumber\\
G &=& \langle {A_i}^\dagger a_j\rangle = \langle {A_i} a_j^\dagger\rangle = \langle {B_i}^\dagger b_j\rangle = \langle {B_i} b_j^\dagger\rangle\nonumber\\ 
g &=& \langle {A_i} B_j\rangle = \langle {A_i}^\dagger {B_j}^\dagger\rangle =  \langle {a_i} b_j\rangle = \langle {a_i}^\dagger {b_j}^\dagger\rangle\nonumber\\
h &=& \langle {A_i} B_j\rangle = \langle {A_i}^\dagger {B_j}^\dagger\rangle =\langle {a_i} b_j\rangle = \langle {a_i}^\dagger {b_j}^\dagger\rangle \ .
\label{eq:parameters}
\end{eqnarray}
The difference between $g$ and $h$ in \eqref{eq:expectation} is that $g$ corresponds to the expectation value of next nearest neighbour pairs in the plane and $h$ corresponds to nearest neighbour pairs between the planes, see Fig.~\ref{fig:magnet}. It is convenient to introduce parameters $\mu$, $\nu$, $\eta$ and $\chi$ defined as
\begin{eqnarray}
\label{eq:quantum}
\mu S &=& S + \frac{1}{2} - f + G\nonumber \\
\nu S &=& S + \frac{1}{2} - f + F\nonumber\\
\eta S &=& S + \frac{1}{2} - f + g\nonumber\\
\chi S &=& S + \frac{1}{2} - f + h \ .
\label{eq:correction}
\end{eqnarray}
Values of the parameters obtained in the self-consistent procedure described below are plotted in Fig.~\ref{fig:quantum}. After the Hartree-Fock decoupling the Hamiltonian \eqref{eq:quarthamilton} is transformed to
\begin{eqnarray}
\label{eq:mfd}
H &=& \alpha + H^\text{AF1} + H^\text{F} + H^\text{AF2}\nonumber\\
\alpha &=& -2J_2S^2 - 2J_cS^2\nonumber\\
H^\text{AF1} &=&  J_1\mu S\sum_{\langle i,j\rangle}
({A_i}^\dagger A_i + {a_j}^\dagger a_j +{B_i}^\dagger B_i + {b_j}^\dagger b_j\nonumber\\
 &-& A_i a_j - {A_i}^\dagger {a_j}^\dagger - B_i b_j - {B_i}^\dagger {b_j}^\dagger)\nonumber\\
H^\text{F} &=& - J_1\nu S\sum_{\langle i,j\rangle}({A_i}^\dagger A_i + {b_j}^\dagger b_j + {a_i}^\dagger a_i + {B_j}^\dagger B_j \nonumber\\
&-& A_i {b_j}^\dagger - {A_i}^\dagger {b_j} - a_i {B_j}^\dagger - {a_i}^\dagger {B_j})\nonumber\\
H^\text{AF2} & =& -J_2\eta S\sum_{\langle\langle i,j\rangle\rangle}({a_i}^\dagger a_i + {b_j}^\dagger b_j + {A_i}^\dagger A_i + {B_j}^\dagger B_j \nonumber\\
&-& a_i {b_j} - {a_i}^\dagger {b_j}^\dagger - A_i {B_j} - {A_i}^\dagger {B_j}^\dagger)\nonumber\\
&+& J_c\chi S\sum_{| i, j|}({a_i}^\dagger a_i + {b_j}^\dagger b_j +{A_i}^\dagger A_i + {B_j}^\dagger B_j - a_i {b_j}\nonumber\\ 
&-& {a_i}^\dagger {b_j}^\dagger - A_i {B_j} - {A_i}^\dagger {B_j}^\dagger) 
\end{eqnarray}
In the momentum representation the Hamiltonian reads
\begin{eqnarray}
\label{eq:FTHamilton}
H &=& \alpha + H^\text{F} + H^\text{AF1} + H^\text{AF2}\nonumber\\
\alpha &=& -2(2J_2 + J_c)NS^2\nonumber\\
H^\text{AF1} &=& 4J_1\mu S\sum_{\bm k\in \text{MBZ}}(({A_k}^\dagger A_k + {a_k}^\dagger a_k + {B_k}^\dagger B_k + {b_k}^\dagger b_k)\nonumber\\
&-&C_{-}(A_k a_{-k} + {A_k}^\dagger {a_{-k}}^\dagger + B_k b_{-k} + {B_k}^\dagger {b_{-k}}^\dagger))\nonumber\\
H^\text{F} &=& - 4J_1\nu S\sum_{\bm k \in \text{MBZ}}(({A_k}^\dagger A_k + {a_k}^\dagger a_k + {B_k}^\dagger B_k + {b_k}^\dagger b_k)\nonumber\\
&-&C_{+}(A_k{b_k}^\dagger + {A_k}^\dagger {b_k} + (a_k{B_k}^\dagger + {a_k}^\dagger {B_k}))\nonumber\\
H^\text{AF2} &=& 4S(2J_2\eta + J_c\chi)\sum_{\bm k \in \text{MBZ}}(({A_k}^\dagger A_k + {a_k}^\dagger a_k \nonumber\\
&+& {B_k}^\dagger B_k + {b_k}^\dagger b_k) - \mu_{\bm k}(A_k {B_{-k}} + {A_k}^\dagger {B_{-k}}^\dagger\nonumber\\ 
&+& a_k {b_{-k}} + {a_k}^\dagger {b_{-k}}^\dagger)) 
\end{eqnarray}

Because of four different sublattices the  Fourier transform is defined in Magnetic Brillouin Zone (MBZ), this is the Brillouin Zone of one sublattice, say the sublattice $a$,
\begin{eqnarray}
\label{eq:fourier}
a_i = \sqrt{\frac{4}{N}}\sum_{\bm k \in \text{MBZ}}e^{-i{\bf k}\cdot{\bf r_i}}a_k  \ .
\end{eqnarray}
In Eq.~(\ref{eq:FTHamilton}) $H^\text{F}$ and $H^\text{AF1}$ consists of ferromagnetic and antiferromagnetic  intraplane terms for nearest neighbour pairs ($J_1$)  while $H^\text{AF2}$ consists  of antiferromagnetic intraplane next nearest neighbour pairs ($J_2$) and antiferromagnetic interplane nearest neighbour pairs ($J_c$). In addition for  simplicity we have introduced the coefficients $C_{+}$, $C_{-}$ and $\mu_{\bm k}$
\begin{eqnarray}
\label{eq:mu}
\mu_{\bm k} &=&\frac{J_2\eta(\cos(k_x) + \cos(k_y)) + J_c\chi \cos(k_z)}{2J_2\eta + J_c\chi}\nonumber\\
C_{+} &=& \cos((k_x+k_y)/2) = \cos(k_b/\sqrt{2})\nonumber\\
C_{-} &=& \cos((k_x - k_y/2) = \cos(k_a/\sqrt{2})\nonumber\\
C_z &=& \cos(k_z) = \cos(k_c)
\end{eqnarray}
The components $k_x$ and $k_y$ are directed along diagonals of the base square, see, Fig.~\ref{fig:magnet}. At this stage it is convenient to unfold the Magnetic Brillouin Zone and to use the full Brillouin Zone.
\begin{eqnarray}
\label{eq:BZ}
{\bf k} = (&k_a&, k_b, k_c)\nonumber\\
-\pi \leq &k_a& \leq \pi\nonumber\\
-\pi \leq &k_b& \leq \pi\nonumber\\
-\pi \leq &k_c& \leq \pi\ 
\end{eqnarray}
Therefore the momentum summation  in all subsequent equations is defined as
\begin{equation}
\label{sBZ}
\sum_{\bm k}= \frac{N}{2}\int_{-\pi}^{\pi}\frac{dk_a}{2\pi}\int_{-\pi}^{\pi}\frac{dk_b}{2\pi}\int_{-\pi}^{\pi}\frac{dk_c}{2\pi} \ .
\end{equation}
Since we are considering the four distinct sublattices $A$, $a$, $B$, and $b$ the full Brillouin zone over counts the number  of degrees of freedom. To compensate this we have introduced an additional prefactor in the definition of summation (\ref{sBZ}).

In the symmetry broken phase the dispersion reads
\begin{equation}
\label{eq:dispersion}
\omega({\bf k}) = \sqrt{{A_{\bm k}}^2 - {B_{\bm k}}^2}
\end{equation}
The coefficients $A_{\bm k}$ and $B_{\bm k}$ are easily obtained by diagonalizing the Hamiltonian \eqref{eq:FTHamilton} in the usual way via a Bogoliubov transformation 
\begin{eqnarray}
\label{eq:coeff}
A_{\bm k} &=& 2J_2S({\lambda + x\nu C_+})\nonumber\\
B_{\bm k} &=&  2J_2S(2\eta C_+C_- + x\mu C_- + y\chi C_z)\ 
\end{eqnarray}
where for simplicity we have introduced the frustration parameters
\begin{eqnarray} 
&&x = J_1/J_2\nonumber\\
&&y = J_c/J_2 \ 
\end{eqnarray}
and have the numerical factor 
\begin{equation}
\lambda = x(\mu - \nu) + 2\eta + y\chi
\end{equation}
which is expressed in terms of the quantum correction parameters \eqref{eq:quantum}. The sublattice magnetisation $m_s$ is defined as $m_s=\langle {S_i}^z\rangle$ so that
\begin{equation}
\label{stag}
m_s = S + \frac{1}{2} -\frac{1}{N}\sum_{\bm k}{\frac{A_{\bm k}}{\sqrt{{A_{\bm k}}^2-{B_{\bm k}}^2}}}
\end{equation}
The quantum correction parameters were determined from the self consistent equations by numerical iteration;  see Eq.~(\ref{eq:expectation}) and Eq.~(\ref{eq:quantum}).
\begin{eqnarray}
\label{eq:FGgh}
f - \frac{1}{2} &=& \frac{1}{N}\sum_{\bm k}{\frac{A_{\bm k}}{\sqrt{{A_{\bm k}^2}-{B_{\bm k}}^2}}}\nonumber\\
F &=& \frac{1}{N}\sum_{\bm k}{\frac{B_{\bm k}C_-}{\sqrt{{A_{\bm k}}^2-{B_{\bm k}}^2}}}\nonumber\\
G &=& \frac{1}{N}\sum_{\bm k}{\frac{A_{\bm k}C_+}{\sqrt{{A_{\bm k}}^2-{B_{\bm k}}^2}}}\nonumber\\
g  &=&  \frac{1}{N}\sum_{\bm k}{\frac{B_{\bm k}\mu_{\bm k}}{\sqrt{{A_{\bm k}}^2-{B_{\bm k}}^2}}}\nonumber\\
h  &=&  \frac{1}{N}\sum_{\bm k}{\frac{B_{\bm k}C_z}{\sqrt{{A_{\bm k}}^2-{B_{\bm k}}^2}}} 
\end{eqnarray}

\begin{figure}

    \begin{center}

     \includegraphics[trim  = 0mm 0mm 0mm 0mm, width=0.85\columnwidth,clip]{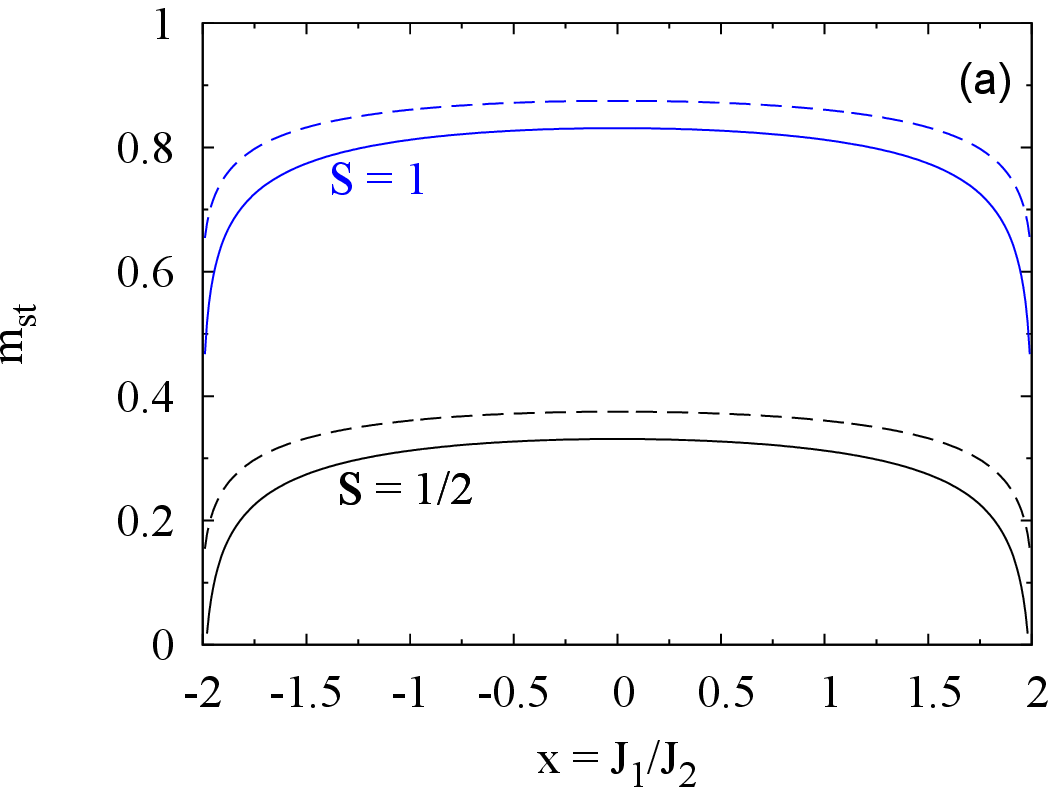}

     \includegraphics[trim  = 0mm 0mm 0mm 0mm, width=0.85\columnwidth,clip]{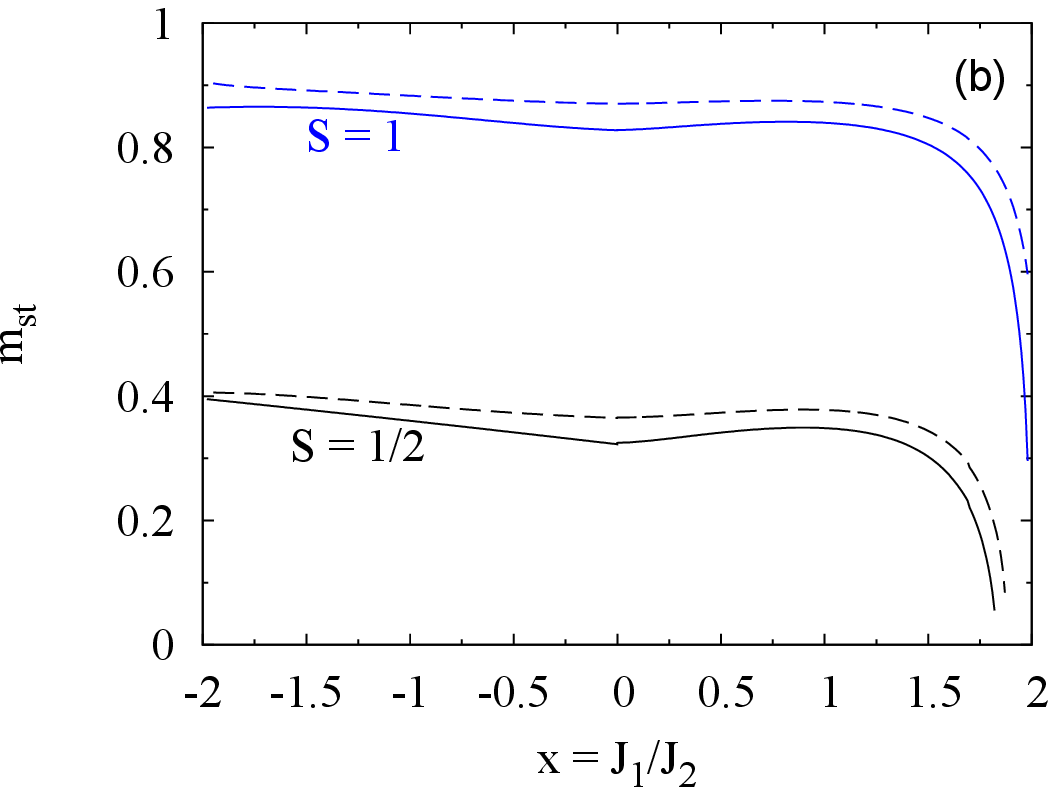}

    \end{center}

    \caption{(colour online) Staggered magnetization in the three-dimensional magnetic Brillouin zone as a function of the ratio of the exchange couplings $J_1$ and $J_2$. The calculated three dimensional staggered magnetisations were obtained using (a) linear spin wave theory and (b) self consistent spin wave theory. In each plot we show results for $S = 1/2$ and $S = 1$ for $y = 0.01$ (solid) and $0.10$ (dashed).}

    \label{fig:staggered}

\end{figure}

\begin{figure}

    \begin{center}

     \includegraphics[trim  = 0mm 0mm  0mm 0mm, width=0.9\columnwidth,clip]{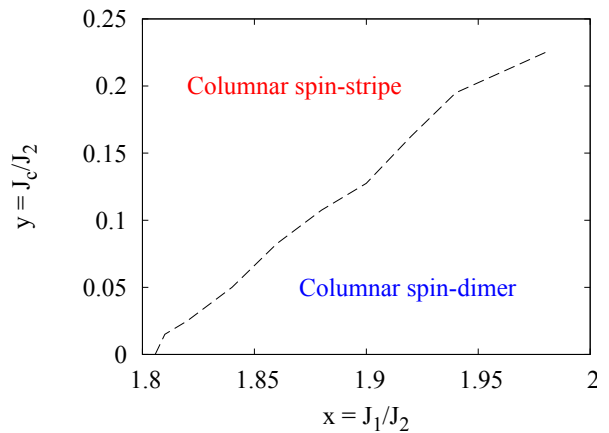}

    \end{center}

    \caption{(colour online) The phase diagram of the $S=1/2$ 3D $J_1$-$J_2$-$J_c$ Heisenberg model, $y = J_c/J_2$  measures the interlayer coupling. The region above the curve corresponds to the columnar spin-stripe phase and the region below to possibly the columnar spin-dimerized phase.}

\label{fig:stable-s12}

\end{figure}

\begin{figure}

    \begin{center}

     \includegraphics[trim  = 0mm 0mm 0mm 0mm, width=0.9\columnwidth,clip]{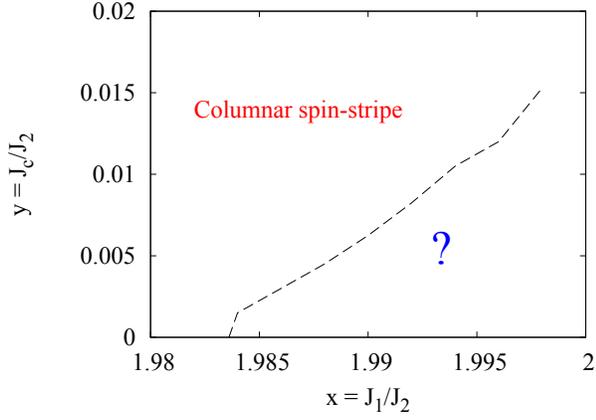}

    \end{center}

    \caption{(colour online) The phase diagram of the $S=1$ 3D $J_1$-$J_2$-$J_c$ Heisenberg model, $y = J_c/J_2$  measures the interlayer coupling. The region above the curve corresponds to the columnar spin-stripe phase and the region below to possibly either a columnar spin-dimerized phase like for $S = 1/2$ or a quadrupoler ordered phase. Note the very different scales compared to Fig.\ \ref{fig:stable-s12}.}

\label{fig:stable-s1}

\end{figure}

For $2J_2$ $>$ $J_1$ $\gg$ $J_c$ $>$ $0$, the classical ground state is the “stripe” ordered phase with ordering vector $(\pi,0,\pi)$. In addition to breaking spin rotational and time-reversal symmetries, the lattice symmetry is spontaneously broken, the ordering vector $(0,\pi,\pi)$ is equally possible. To be specific we consider the $(\pi,0,\pi)$ phase. The excitation spectrum is gapless at $(0,0,0)$ and $(\pi,0,\pi)$. These are the Goldstone modes with dispersion
\begin{equation}
\label{eq:dispersion2}
\omega({\bf q}) \approx \sqrt{{v_a}^2 {q_a}^2 + {v_b}^2 {q_b}^2 + {v_c}^2 {q_c}^2}
\end{equation}
where $q$ is measured relative to the ordering vector and $v_a$, $v_b$, $v_c$ are the spin wave velocities along the crystal axes
\begin{eqnarray}
\label{eq:velocities}
v_a &=& 2J_2S\sqrt{(2\eta + x\mu)(2\eta + x\mu + y\chi)}\nonumber\\
v_b &=& 2J_2S\sqrt{(2\eta - x\nu)(2\eta + x\mu + y\chi)}\nonumber\\
v_c &=& 2J_2S\sqrt{y\chi(2\eta + x\mu + y\chi)} 
\end{eqnarray}
The existence of this branch of magnetic excitations follows from the Goldstone theorem. There exists also a second branch of spin waves for $\omega(0,\pi,0) = \omega(\pi,\pi,\pi)$ and $\omega(0,\pi,\pi) = \omega(\pi,\pi,0)$ which have not yet been observed experimentally. Since they are gapped the intensity of scattered neutrons would be substantially lower than the primary branch at $(\pi,0,\pi)$. Their energies read
\begin{eqnarray}
\label{eq:primarybranch}
\omega(0,\pi,0) &=& 4J_2S\sqrt{(2\eta-x\nu)(x\mu-x\nu+y\chi)}\nonumber\\
\omega(0,\pi,\pi) &=& 4J_2S\sqrt{(x\mu-x\nu)(2\eta -x\nu +y\chi)}
\end{eqnarray}

\begin{figure}

    \begin{center}

     \includegraphics[trim  = 10mm 5mm 10mm 0mm, width=0.85\columnwidth,clip]{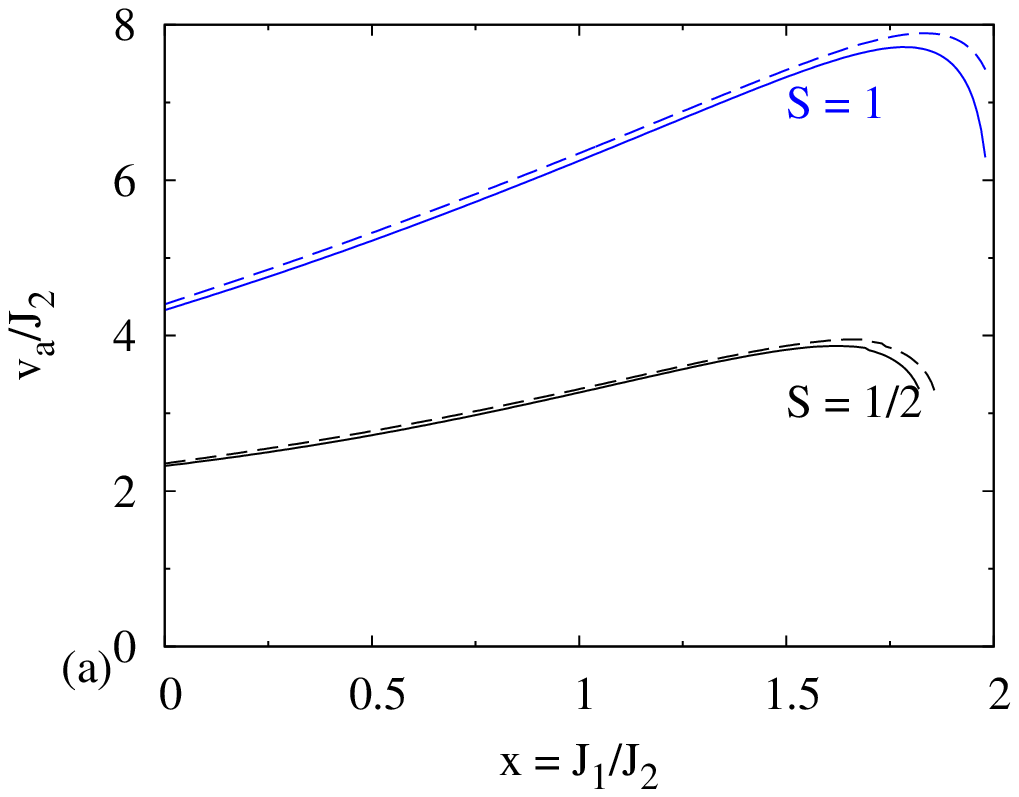}

     \includegraphics[trim  = 10mm 5mm 10mm 0mm, width=0.85\columnwidth,clip]{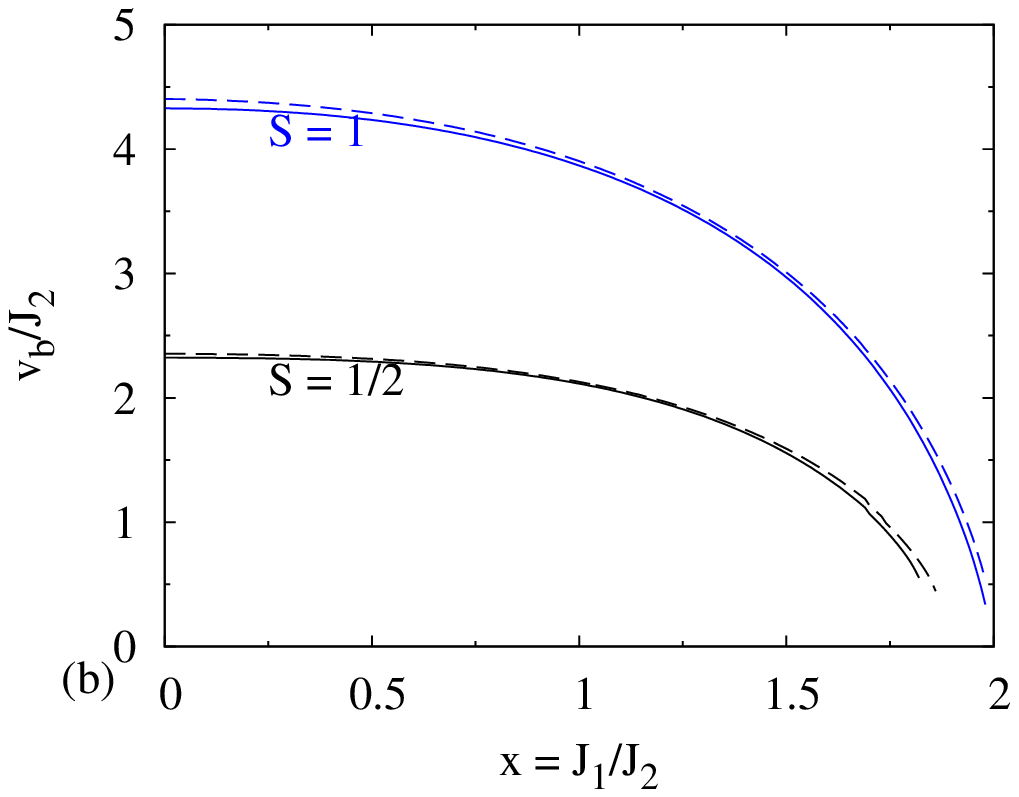}

     \includegraphics[trim  = 10mm 5mm 10mm 0mm, width=0.85\columnwidth,clip]{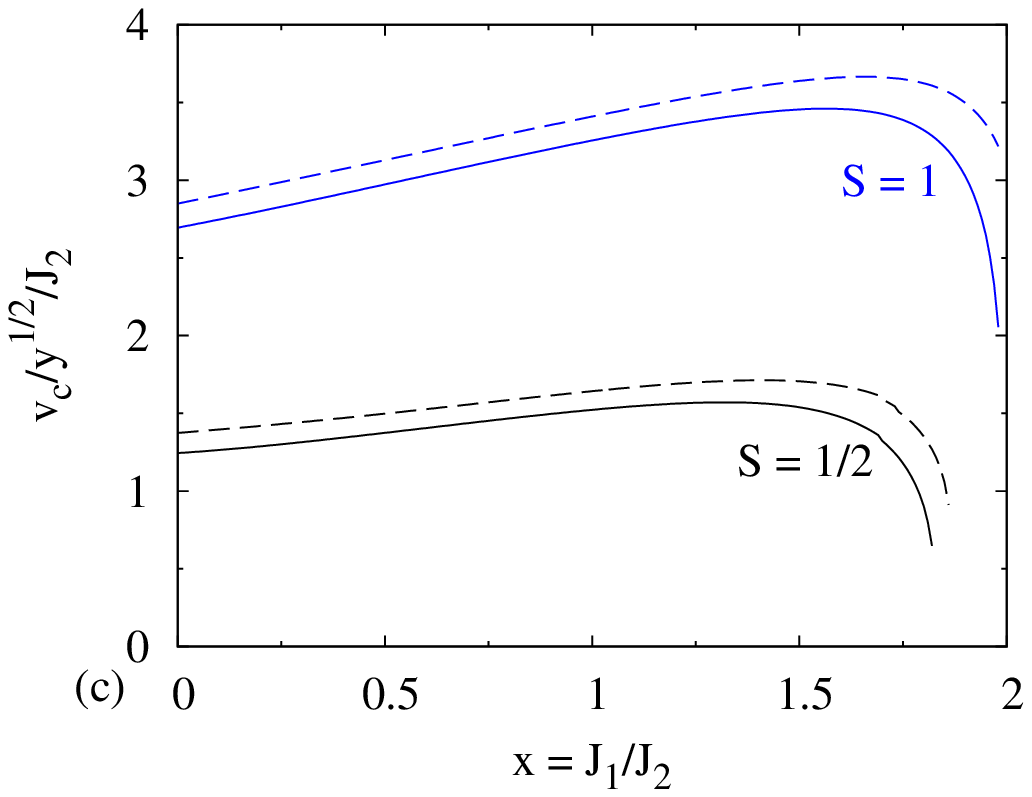}

    \end{center}

    \caption{(colour online) Spin wave velocities along the crystal axes as functions of the ratio $J_1/J_2$  for $y = 0.01$ (solid) and $0.10$ (dashed). Results are presented for $S =1/2$ and $S=1$. Note that the ordinate values are divided by $\sqrt{y}$ for $v_{c}$ only.}

\label{fig:velocities}

\end{figure}

\section{General Results}

\label{sec:general}

The influence of the ratio of exchange parameters $y = J_c/J_2$ on the resulting staggered magnetization, spin wave velocities and excitation spectra is studied for a range of values of $y$. The staggered magnetisation was calculated by numerical iteration for a range of $y$ assuming $y \ll 1$; the results are shown in Fig.~\ref{fig:staggered}. We show results for both linear spin wave theory and self -consistent spin wave theory for $S =1/2$ and $S = 1$.

We have also calculated the ``critical'' value of $x=J_1/J_2$ where either the staggered magnetization or one of the spin wave velocities vanishes, both, for  $S=1/2$ in Fig.~\ref{fig:stable-s12} and for $S=1$ in Fig.~\ref{fig:stable-s1}. A detailed distinction between vanishing staggered magnetization or vanishing spin wave velocity will be presented elsewhere \cite{stane10}. Basically these figures show the phase diagram of the model as it results from self-consistent spin wave theory. For $S=1/2$, the regions above the lines correspond to the columnar spin-stripe phase and the regions below them correspond to the columnar spin-dimerized phase. It is well known that the transition between these phases is of first order. Therefore, strictly speaking, the criterion of a vanishing magnetization is not quite the correct one to indicate the transition, the true transition happens at a slightly smaller value of $J_1/J_2$ than that indicated in  Fig.~\ref{fig:stable-s12}. However, it is known, see e.g. Ref.~\onlinecite{singh99b}, that the criterion gives practically the correct value of the critical point for $y = 0$. Here we assume that the same is true for small nonzero $y$.

In Fig.~\ref{fig:staggered} we observe an interesting behaviour of $m_s (x)$ as a function of increasing interlayer coupling. In the vicinity of $x=2$ and for small values of $y$ the renormalized magnetization changes very rapidly on small parameter changes, differing significantly from linear spin wave theory calculations even for $S=1$. This clearly indicates strong quantum fluctuations. These fluctuations are stronger for $S=1/2$, see for instance the larger deviation of the region of instability from the value $x=2$, than for $S=1$ (compare Fig.~\ref{fig:stable-s12} and Fig.~\ref{fig:stable-s1}). This also implies that very small values of the staggered magnetization can be obtained for $S=1/2$ more easily, i.e., with less fine-tuning, than for $S=1$. But qualitatively, the curves for $S=1/2$ and for $S=1$ are very similar. For all parameters, the long-range order is strengthened by the coupling in the third dimension.

We would like to briefly comment on the intermediate quantum phases for $S = 1/2$ and $S = 1$ near $x = 2$. There is a consensus that in the $S = 1/2$ two-dimensional $J_1$-$J_2$ model there is an intermediate magnetically disordered phase at $0.4J_1 < J_2 < 0.6J_1$. We believe that this is the columnar spin dimer-phase. According to our data for the 3D model presented in Fig.~\ref{fig:stable-s12} the columnar spin dimer phase disappears around $J_{c}/J_{2} \approx 0.25$. The critical value of $J_c$ for disappearance of the magnetically disordered phase differs from that determined previously \cite{schma06} by the couple-cluster method, $J_{2}/J_{1} \approx 0.36$. We do not think that the difference in the value is significant, both our method and the method of Ref.~\onlinecite{schma06} are approximate. More importantly, there is qualitative agreement about the phase diagram. For $S = 1$ we also found a tiny region of an intermediate nonmagnetic phase shown in  Fig.~\ref{fig:stable-s1}. This is qualitatively different from previous studies~\cite{bishop08} that were unable to identify an intermediate phase. Note the very small scale on which we find the instability of the columnar stripe order. Unfortunately within the present method we cannot determine the exact nature of the phase and therefore in Fig.~\ref{fig:stable-s1}  the phase is shown by a question mark.

The ratio of the spin wave velocities along the different directions is very sensitive to the value $x=J_1/J_2$. The point is that in a real compound the staggered magnetization can depend on a range of additional uncontrolled variables such as itinerancy, hybridization, etc. The dispersion relation however only depends on the effective Hamiltonian and thus is less ambiguous \cite{uhrig09a}. Our results for the spin wave velocities along the three crystal axes are shown in Fig.~\ref{fig:velocities} versus the ratio $J_1/J_2$ for $y = 0.01$, and $0.10$. The velocities are given in units of $J_2$; $v_a$ and $v_b$ only weakly depend on $y$, while on the other hand $v_c\propto \sqrt{y}$. The dependence of the ratios of the spin wave velocities on the values of the exchange couplings is stronger than the corresponding dependence of the staggered  magnetization. So it is more appropriate to determine the values of the couplings from the spin wave velocities.

\begin{figure}

    \begin{center}

     \includegraphics[trim  = 10mm 10mm 15mm 15mm,width=1.00\columnwidth,clip]{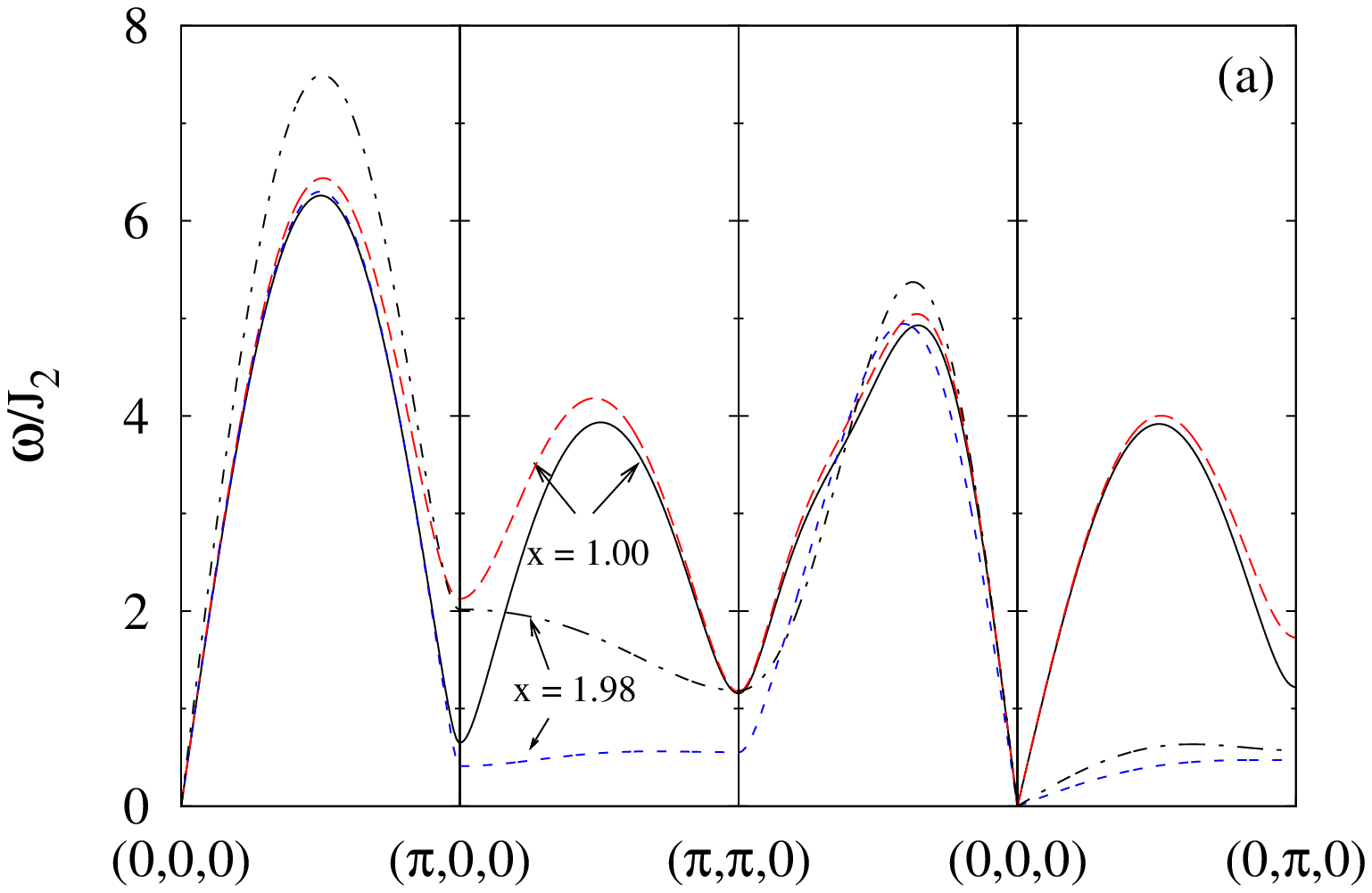}

     \includegraphics[trim  = 10mm 10mm 15mm 15mm,width=1.00\columnwidth,clip]{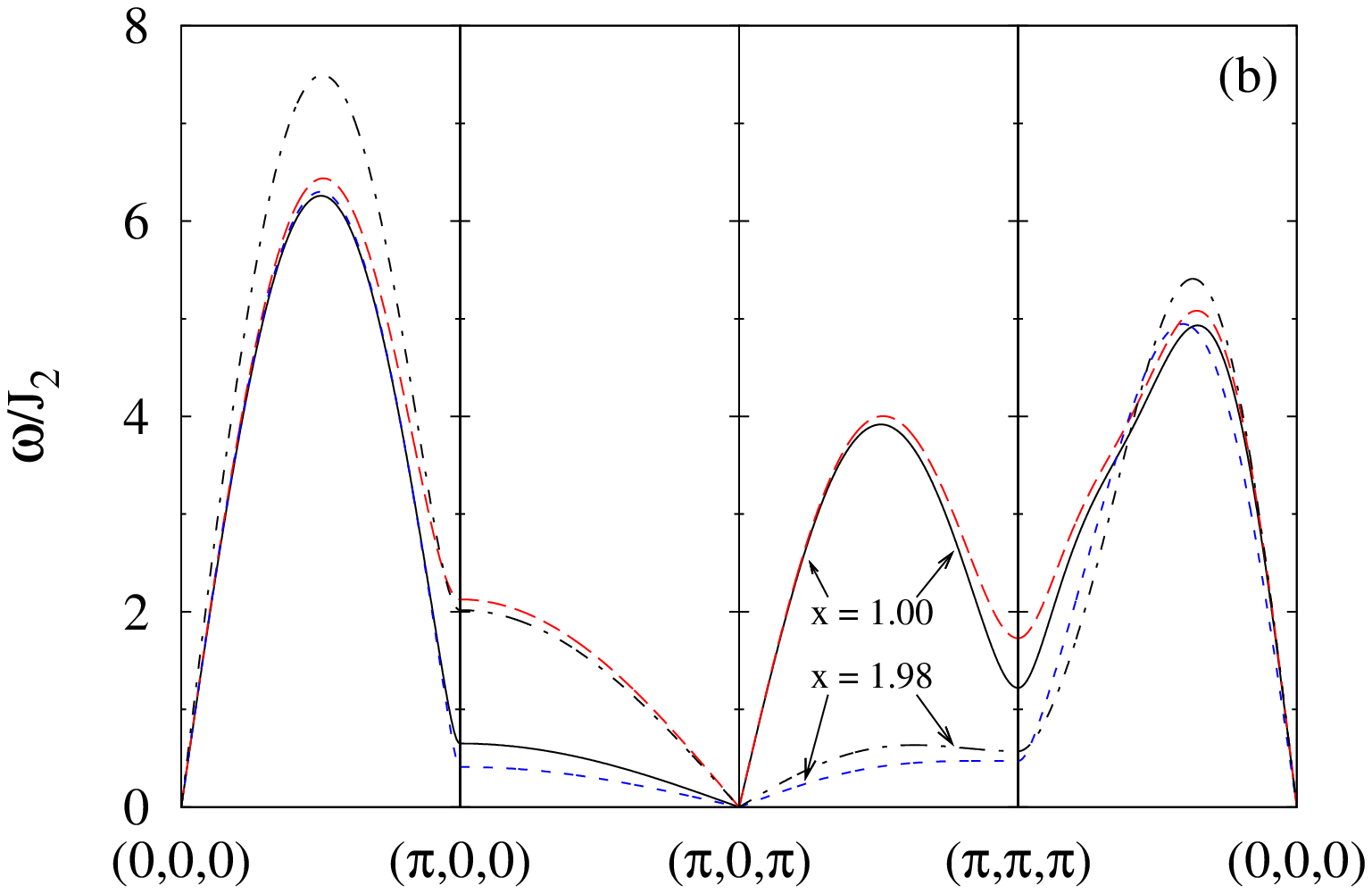}

    \end{center}

    \caption{(colour online) Spin excitation spectra along high-symmetry cuts through the Brillouin zone for $S = 1$. We show the excitation spectra for both a system deep in the columnar phase ($J_1/J_2 = 1$) and one near the quantum phase transition ($J_1/J_2 = 1.98$). Calculations were performed using self-consistent spin wave theory and here we compare results for $y = 0.01$ (solid and dotted) and $0.10$ (dashed and dot-dashed).}

\label{fig:spectra}

\end{figure}

Spin excitation spectra for $S=1$ along high-symmetry cuts through the Brillouin zone are shown in Fig.~\ref{fig:spectra}. We show the excitation spectra for both a system deep in the columnar phase ($J_1/J_2 = 1$) and one near the quantum phase transition ($J_1/J_2 = 1.98$). In the present work we analyze the dependence of the spin wave dispersion on the interlayer coupling $J_c$. The case of the small coupling, $y=J_c/J_2 \ll 1$ is of special interest. Expanding the spectrum as defined in \eqref{eq:dispersion} in powers of $y$ we find  
\begin{equation}
\label{eq:fitting}
\omega^2({\bf k}) = \omega_0^2({\bf k}) +y\ \delta\omega^2({\bf k})
\end{equation}
where  
\begin{eqnarray}
\label{eq:2ddispersion}
\omega_0^2(\bf k) &=& 4J_2^2S^2[(x(\mu -\nu) + 2\eta + x\nu C_{+})^2\nonumber\\ 
&-& (2\eta C_{+} C_{-} + x\mu C_{-})^2] 
\end{eqnarray}
and
\begin{eqnarray}
\label{eq:fitting2}
\delta\omega^2(\bf k) &=& 8J_2^2S^2\chi[(x(\mu -\nu) + 2\eta + x\nu C_{+})\nonumber\\ 
&-& (2\eta C_{+} C_{-} + x\mu C_{-})C_z] 
\end{eqnarray}
Note that all parameters in Eq.~(\ref{eq:2ddispersion}) and Eq.~(\ref{eq:fitting2}) are calculated at $y=0$. Eq.~(\ref{eq:2ddispersion}) is the dispersion in the 2D case, it has Goldstone modes $(k_a,k_b)=(0,0)$ and $(k_a,k_b)=(\pi,0)$. The expanded 3D dispersion \eqref{eq:fitting}  has the Goldstone modes at $(k_a,k_b,k_c)=(0,0,0)$ and $(k_a,k_b,k_c)=(\pi,0,\pi)$ as expected. Note that $\omega({\bf k})$ is generally a non-analytic function of $y$ at small $y$ while $\omega^2({\bf k})$ is the analytic one. This is why the expansion \eqref{eq:fitting} is written in terms of $\omega^2$. In the limit $y\to 0$ the parameters $\mu,\nu,\eta,\chi$ depend on $x=J_1/J_2$ only. The corresponding plots are presented in Fig.~\ref{fig:quantum}. These plots together with Eq.~(\ref{eq:fitting}) allow one to determine the spin-wave dispersion at arbitrary small $y$.

\section{Application to Iron Pnictides}

\label{sec:iron}

\begin{table}

\begin{center}

    \begin{tabular}{ | c | c | c | }

\hline

    Material & LDA Moment \cite{han09} ($\mu_B) $ & Expt. Moment ($\mu_B$)

    \\ \hline

   LaOFeAs & 1.69 & 0.36 \cite{cruz08} \\ \hline

   NdOFeAs & 1.49 & 0.25 \cite{chen08} \\ \hline

   CaFe$_2$As$_2$ & 1.51 & 0.80 \cite{diall09}\\ \hline

   BaFe$_2$As$_2$ & 1.68 & 0.87 \cite{huang08} \\ \hline

   SrFe$_2$As$_2$ & 1.69 & 1.01 \cite{kanek08}\\ \hline	

    \end{tabular}

\end{center}

  \caption{Comparison of the magnetic moment in units of $\mu_B$ as predicted from LDA calculations and those observed in experiment.}

  \label{moment}

\end{table}

The smallness of the measured magnetic moment in the iron pnictides relative to theoretical calculations is a matter of controversy; there exist two different scenarios which offer different explanations for the discrepancy between experiment and theory. It has been suggested that magnetic fluctuations may strongly reduce the local magnetic moment, with the ratio $x = J_1/J_2$ being fixed to an appropriate value in the critical scenario. The alternative derives its explanation from the role of the local electronic orbitals and therefore the magnetic couplings are not determined by the value of the magnetic moment. Band structure calculations have shown that $J_1$ and $J_2$ are antiferromagnetic and very similar in value \cite{ma09, yin08, cao08}. In Table \ref{moment} we compare the size of the experimentally measured magnetic moment to that calculated by LDA methods. Typically the experimental values are at least twice smaller than the LDA values. Due to strong quantum fluctuations in the model in principle one can obtain the required suppression of the staggered magnetization by two times by choosing $x\approx 1.99$ for $S = 1$, see Fig.~\ref{fig:staggered}(b). However, due to the large change in $m_s$ upon small parameter changes, it is clear from Fig.~\ref{fig:staggered}(b) that considerable fine tuning is required.

\begin{figure}

    \begin{center}

     \includegraphics[width=0.9\columnwidth,clip]{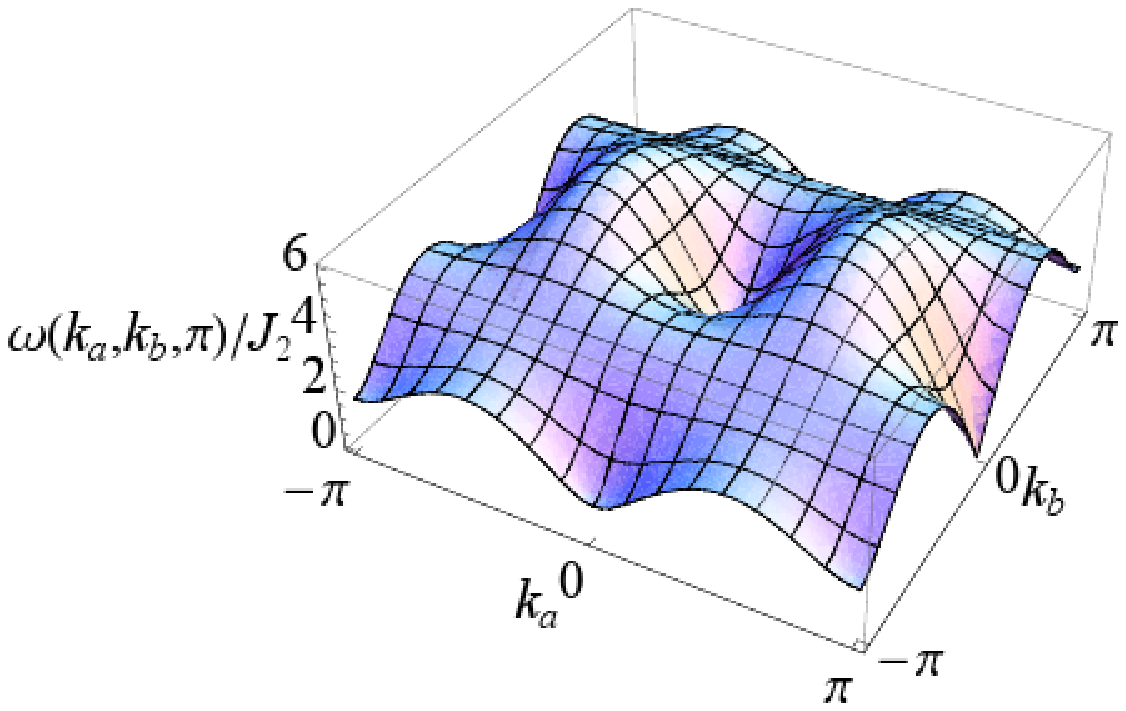}

     \includegraphics[width=0.9\columnwidth,clip]{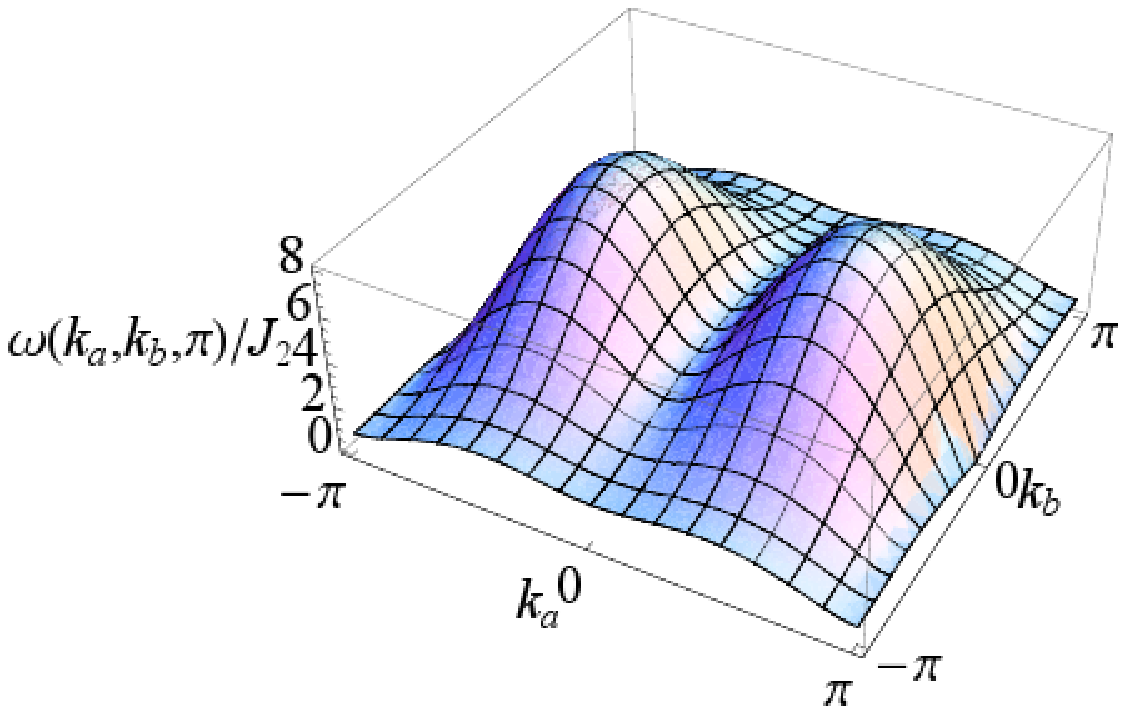}

    \end{center}

    \caption{(colour online) Spin wave dispersion in the plane of the spin stripes. The spin wave dispersions $\omega({\bf k})$ at $\omega(k_a, k_b,\pi)$ in units of $J_2$ is shown for $J_1/J_2 = 0.76$ (top) and $J_1/J_2 = 1.972$ (bottom) for $y = 0.10$. Values were obtained using self consistent spin wave theory for $S = 1$.}

    \label{fig:planespectra}

\end{figure}

\begin{figure}

    \begin{center}

     \includegraphics[width=0.9\columnwidth,clip]{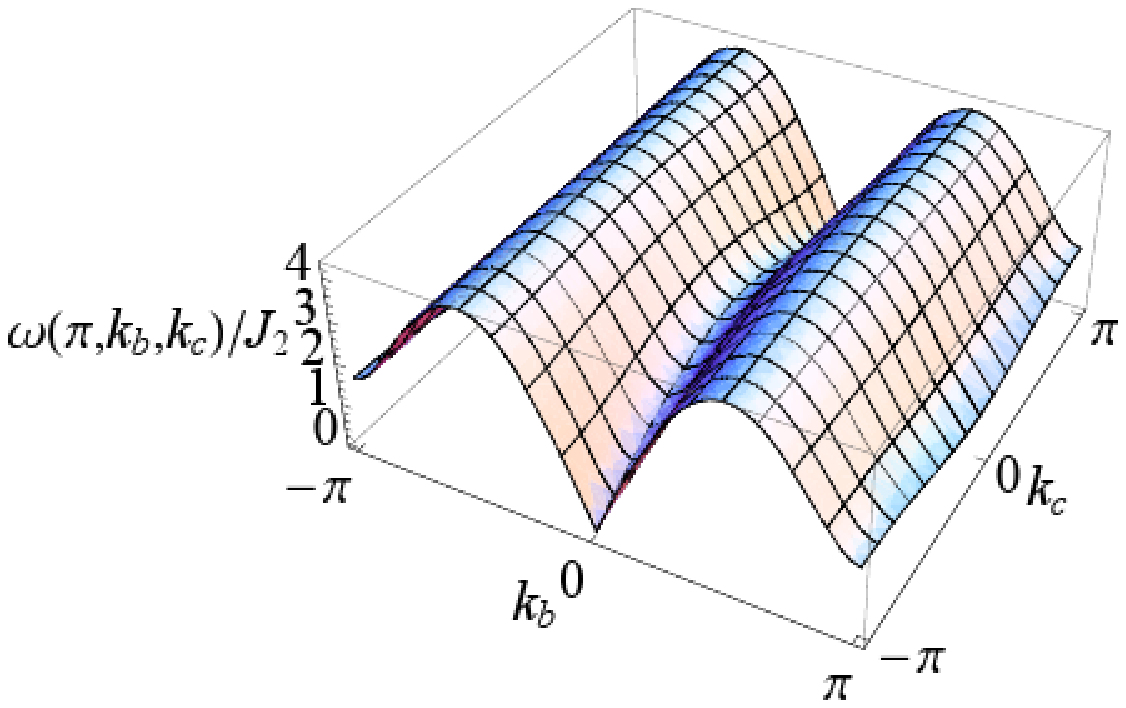}

     \includegraphics[width=0.9\columnwidth,clip]{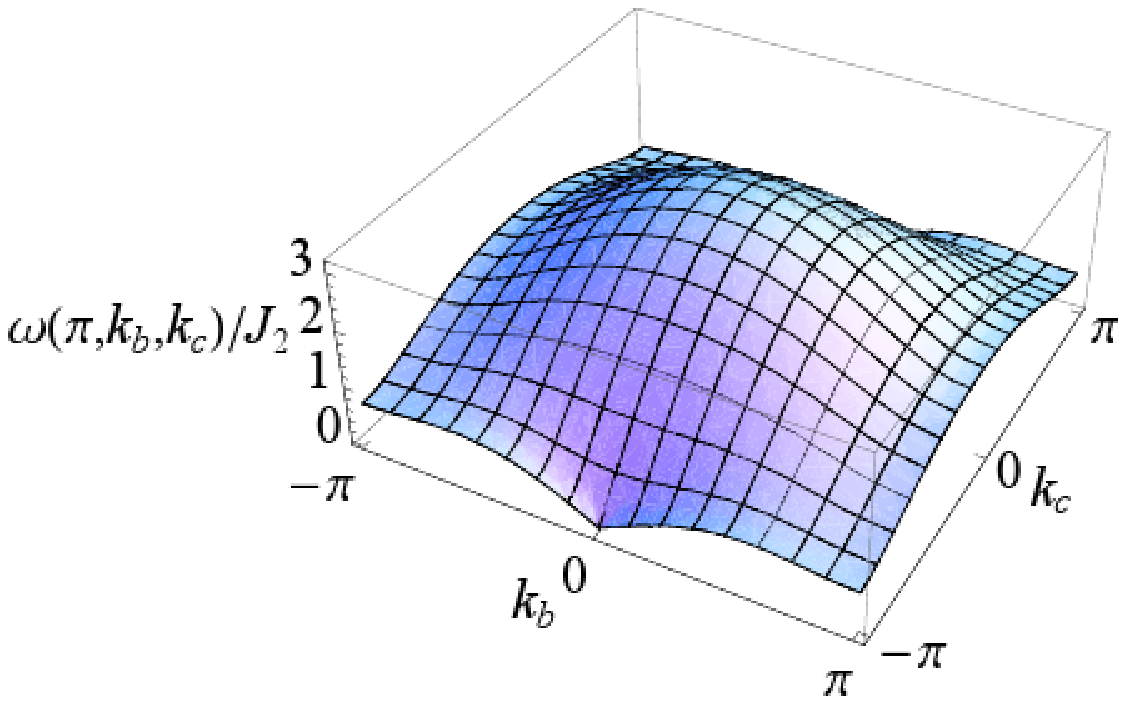}

    \end{center}

    \caption{(colour online) Spin wave dispersion in the plane along and through the spin stripes. The spin wave dispersions  $\omega({\bf k})$ at $\omega(\pi, k_b, k_c)$ in units of $J_2$ is shown for $J_1/J_2 = 0.76$ (top) and $J_1/J_2 = 1.972$  (bottom) for $y = 0.10$.  Values were obtained using self consistent spin wave theory for $S = 1$.}

    \label{fig:interplanespectra}

\end{figure}

\begin{figure}

    \begin{center}

     \includegraphics[trim  = 0mm 0mm 0mm 0mm, width=0.9\columnwidth,clip]{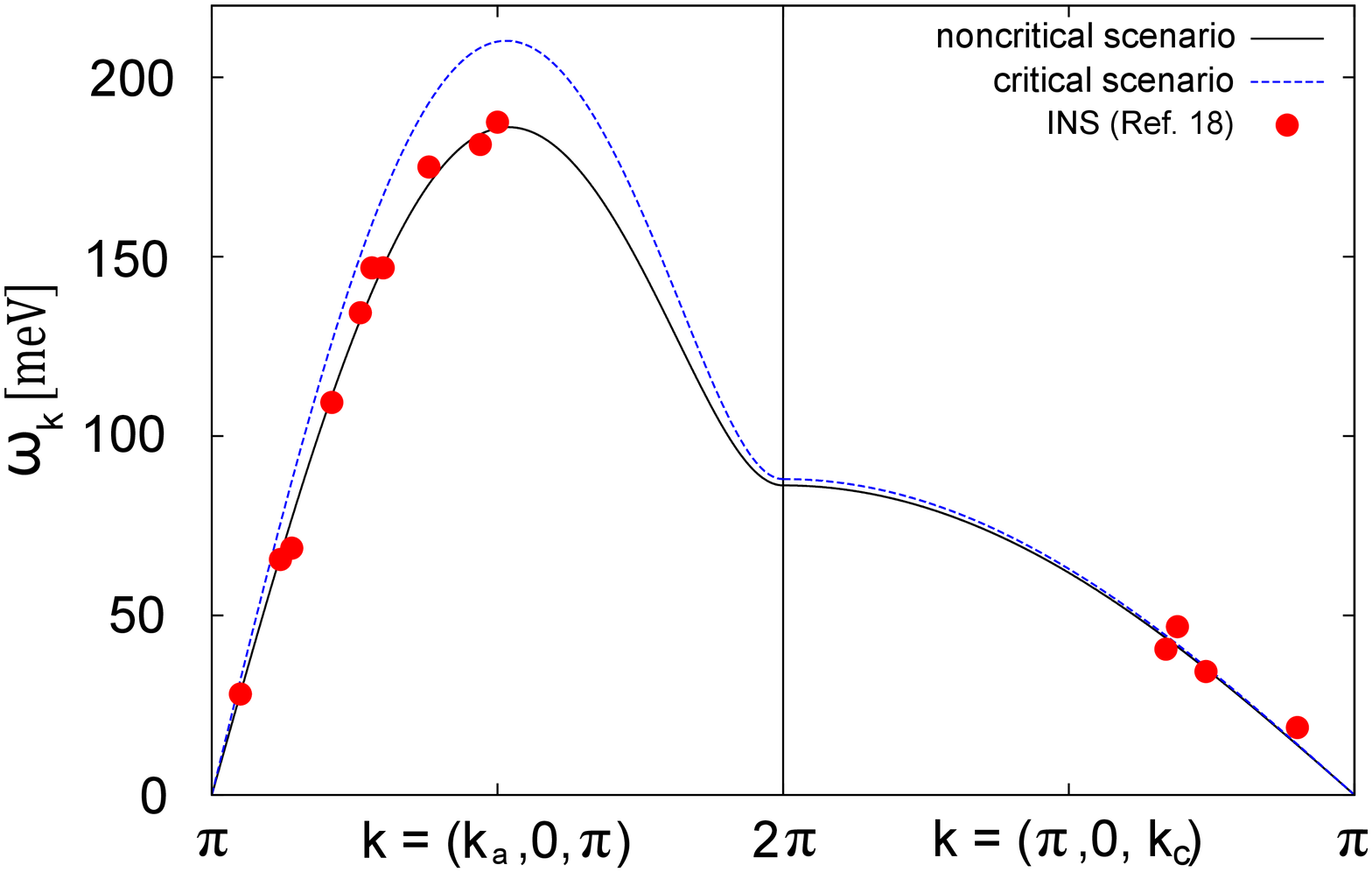}

     \includegraphics[trim  = 0mm 0mm 0mm 0mm, width=0.9\columnwidth,clip]{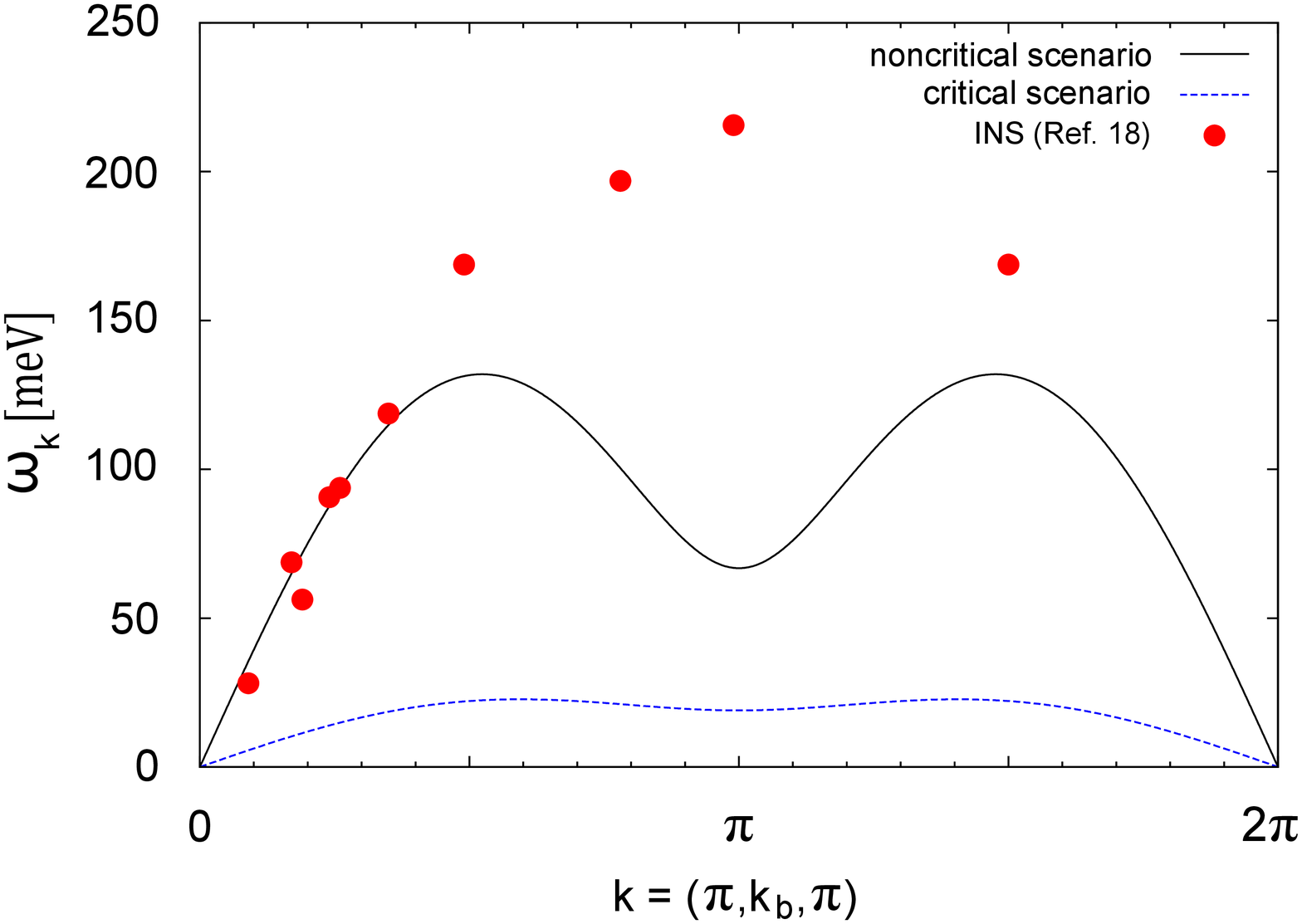}

    \end{center}

    \caption{(colour online) Comparison of fitted dispersions for the critical and noncritical scenarios with inelastic neutron scattering data. We see that the noncritical scenario agrees nicely with the experimental data whereas the critcal scenario does not describe the disperison at higher energies due to problems with the reduced staggered moment and the anisotropic spin wave velocities.}

    \label{fig:zhaodata}

\end{figure}

A more effective probe for the value of the ratio $x = J_1/J_2$ is considering the spin wave velocities since they depend on the Hamiltonian only. In the present work we compare the $J_1$-$J_2$-$J_c$ Heisenberg model to experimental evidence; this is studied via the critical and non-critical scenarios. The dispersion in the plane of the spin stripes $\omega(k_a, k_b, \pi)$ is shown in  Fig.~\ref{fig:planespectra}. For the critical scenario we adjusted the ratio $x = J_1/J_2=1.974$ to fit the ratios $v_b/v_a$  and $v_c/v_a$ obtained from the analysis~\cite{ong09} of the NMR relaxation rate. For the non-critical scenario we used the experimental ratios of the  spin wave velocities $v_b/v_a \approx 0.70$ and $v_c/v_a \approx 0.25$ that follow from inelastic neutron scattering data~\cite{zhao09}. In this case $x = 0.76$ gives the best fit. In addition we plot in Fig.~\ref{fig:interplanespectra} the dispersion of spin waves in the plane along and through the spin stripes $\omega(\pi ,k_b,k_c )$. The dispersion $\omega(\pi,k_b,k_c )$ is helpful since it shows the effect of the interlayer coupling and how extensive the spin waves propagate along the $c$-axes. In addition it is a way to determine if strong quantum fluctuations persist in three dimensions.

The plots in Fig.~\ref{fig:planespectra} and Fig.~\ref{fig:interplanespectra} provide a general qualitative overview. A quantitative comparison between the critical and noncritical scenarios and the inelastic neutron scattering data is depicted in  Fig.~\ref{fig:zhaodata}. We compare the fitted dispersions for the critical and noncritical scenarios. The noncritical scenario agrees nicely with the experimental data whereas the critcal scenario does not describe the dispersion at higher energies due to problems with the reduced staggered moment and the anisotropic spin wave velocities. Therefore, it is quite clear that the pnictides, if decribed by the $J_1$-$J_2$-$J_c$ Heisenberg model, are deep in the columnar phase with $J_1 = 0.76J_2$.

In addition to determining if the pnictides are in the critical or non-critical regime we study whether or not the $J_1$-$J_2$-$J_c$ Heisenberg model is an appropriate model for describing the undoped pnictides. In the critical scenario, $v_b$ is small compared to $v_a$ implying a high density of magnetic excitations \cite{ong09} while in the non-critical scenario $v_b$ is sizeable compared to $v_a$ implying a low density of magnetic excitations. In Fig.~\ref{fig:zhaodata} we observe that while the noncritical scenario does reproduce the known spin wave velocities, the dispersion cannot be matched globally. Significant differences persist at intermediate energies for the $(\pi, k_b, \pi)$ dispersion curve.  Therefore, it is clear that the $J_1$-$J_2$-$J_c$ Heisenberg model is not consistent with the  data by Zhao et al.\ \cite{zhao09}. We stress that the data is not consistent with the ``isotropic'' Heisenberg model considered in the present work because of the minute orthorhombic distortion. On the other hand, the data is consistent with the anisotropic Heisenberg model, see discussion in Ref.~online\cite{zhao09}. The anisotropic Heisenberg model implies that there are additional degrees of freedom (orbital?) and it is outside of the scope of the present work.

It has been recently suggested that the strong reduction of the magnetic moment possible in two dimensions \cite{uhrig09a} is not possible for substantial three dimensional coupling and/or magnetic anisotropy because these additional couplings dramatically suppress the quantum corrections to the ordered moment \cite{smera10}. The logarithmic  divergence of the quantum corrections to the staggered magnetisation seen in the square lattice $J_1$-$J_2$ model for $x \to x_c$ is cut off by the addition of either an anisotropy gap or a third dimension. In  Fig.~\ref{fig:staggered} we do indeed see an enhancement of the staggered magnetisation compared to the two dimensional case \cite{uhrig09a} for increasing values of $y$ for $S=1/2$ and $S=1$.

Since the quantum fluctuations are more significant for $S=1/2$ than for $S=1$ their suppression by the three dimensional coupling is seen more distinctly in the $S=1/2$ data than in the $S=1$ data. In particular, close to the critical values $x_c$, where the columnar striped phase becomes unstable due to quantum fluctuations, the suppression for increasing three-dimensional coupling $y\to 1$ is most clearly seen, cf.  Fig.~\ref{fig:staggered}, which is in accordance with the findings by Smerald and Shannon \cite{smera10}. The quantitative difference between the results for $S=1/2$ and for $S=1$ is most strikingly seen in the difference of scales of the Figs.~\ref{fig:stable-s12} and \ref{fig:stable-s1}.

\section{Conclusions}

\label{sec:conclusion}

We have provided a quantitative theory for the magnetic excitations on a tetragonal lattice based on a minimal spin model, namely the $J_1$-$J_2$-$J_c$ Heisenberg model. First, we have shown that the columnar phase is stabilized by the introduction of the interlayer coupling $J_c$. Such a three-dimensional coupling dramatically strengthens the staggered magnetization and suppresses the strong quantum fluctuations. Since for the $S = 1/2$ case the staggered magnetization is more strongly renormalized by the quantum fluctuations than for $S=1$ the effects of the suppression of the quantum fluctuations are more clearly seen for $S=1/2$ than for $S=1$. Yet both spin species behave qualitatively the same

In  addition, we have also shown that the position of the critical point depends on the value of of the relative interlayer coupling $y$. Again, this dependence is more significant for $S=1/2$ than for $S=1$ because the influence of quantum fluctuations and thus also of their suppression is stronger for smaller spin than for larger spin. As function of $y$ we found to distinct ways how the columnar phase becomes unstable. Either the magnetization or one of the spin wave velocities vanishes.

Second, the strong reduction of the magnetic moment is possible in three dimensions if one considers  small values of $y$. One has to approach $x_c$ very closely even for $S=1/2$. Third, we have shown that one can conveniently model $\omega_k (y)$ for small $y$ analytically.

Finally, comparing the $J_1$-$J_2$-$J_c$  Heisenberg model with experimental data we found that such a model does not explain the data \cite{zhao09}. If reproduces the spin wave velocities, it cannot match the dispersion globally. It is clear that further work is called for.


\begin{thebibliography}{43}

\expandafter\ifx\csname natexlab\endcsname\relax\def\natexlab#1{#1}\fi

\expandafter\ifx\csname bibnamefont\endcsname\relax

  \def\bibnamefont#1{#1}\fi

\expandafter\ifx\csname bibfnamefont\endcsname\relax

  \def\bibfnamefont#1{#1}\fi

\expandafter\ifx\csname citenamefont\endcsname\relax

  \def\citenamefont#1{#1}\fi

\expandafter\ifx\csname url\endcsname\relax

  \def\url#1{\texttt{#1}}\fi

\expandafter\ifx\csname urlprefix\endcsname\relax\def\urlprefix{URL }\fi

\providecommand{\bibinfo}[2]{#2}

\providecommand{\eprint}[2][]{\url{#2}}



\bibitem[{\citenamefont{Betts and Oitmaa}(1977)}]{betts77}
\bibinfo{author}{\bibfnamefont{D.~D.} \bibnamefont{Betts}} \bibnamefont{and}
  \bibinfo{author}{\bibfnamefont{J.}~\bibnamefont{Oitmaa}},
  \bibinfo{journal}{Phys. Lett.} \textbf{\bibinfo{volume}{62A}},
  \bibinfo{pages}{277} (\bibinfo{year}{1977}).


\bibitem[{\citenamefont{Manousakis}(1991)}]{manou91}
\bibinfo{author}{\bibfnamefont{E.}~\bibnamefont{Manousakis}},
  \bibinfo{journal}{Rev. Mod. Phys.} \textbf{\bibinfo{volume}{63}},
  \bibinfo{pages}{1} (\bibinfo{year}{1991}).


\bibitem[{\citenamefont{Singh et~al.}(1999)\citenamefont{Singh, Weihong, Hamer,and Oitmaa}}]{singh99b}
\bibinfo{author}{\bibfnamefont{R.~R.~P.} \bibnamefont{Singh}},
  \bibinfo{author}{\bibfnamefont{Z.}~\bibnamefont{Weihong}},
  \bibinfo{author}{\bibfnamefont{C.~J.} \bibnamefont{Hamer}}, \bibnamefont{and}
  \bibinfo{author}{\bibfnamefont{J.}~\bibnamefont{Oitmaa}},
  \bibinfo{journal}{Phys. Rev. B} \textbf{\bibinfo{volume}{60}},
  \bibinfo{pages}{7278} (\bibinfo{year}{1999}).


\bibitem[{\citenamefont{Kotov et~al.}(1999)\citenamefont{Kotov, Oitmaa, Sushkov, and Weihong}}]{kotov99b}
\bibinfo{author}{\bibfnamefont{V.~N.} \bibnamefont{Kotov}},
  \bibinfo{author}{\bibfnamefont{J.}~\bibnamefont{Oitmaa}},
  \bibinfo{author}{\bibfnamefont{O.~P.} \bibnamefont{Sushkov}},
  \bibnamefont{and} \bibinfo{author}{\bibfnamefont{Z.}~\bibnamefont{Weihong}},
  \bibinfo{journal}{Phys. Rev. B} \textbf{\bibinfo{volume}{60}},
  \bibinfo{pages}{14613} (\bibinfo{year}{1999}).


\bibitem[{\citenamefont{Sushkov et~al.}(2001)\citenamefont{Sushkov, Oitmaa, and Weihong}}]{sushk01}
\bibinfo{author}{\bibfnamefont{O.~P.} \bibnamefont{Sushkov}},
  \bibinfo{author}{\bibfnamefont{J.}~\bibnamefont{Oitmaa}}, \bibnamefont{and}
  \bibinfo{author}{\bibfnamefont{Z.}~\bibnamefont{Weihong}},
  \bibinfo{journal}{Phys. Rev. B} \textbf{\bibinfo{volume}{63}},
  \bibinfo{pages}{104420} (\bibinfo{year}{2001}).


\bibitem[{\citenamefont{Singh et~al.}(2003)\citenamefont{Singh, Zheng, Oitmaa,Sushkov, and Hamer}}]{singh03}
\bibinfo{author}{\bibfnamefont{R.~R.~P.} \bibnamefont{Singh}},
  \bibinfo{author}{\bibfnamefont{W.}~\bibnamefont{Zheng}},
  \bibinfo{author}{\bibfnamefont{J.}~\bibnamefont{Oitmaa}},
  \bibinfo{author}{\bibfnamefont{O.~P.} \bibnamefont{Sushkov}},
  \bibnamefont{and} \bibinfo{author}{\bibfnamefont{C.~J.} \bibnamefont{Hamer}},
  \bibinfo{journal}{Phys. Rev. Lett.} \textbf{\bibinfo{volume}{91}},
  \bibinfo{pages}{017201} (\bibinfo{year}{2003}).


\bibitem[{\citenamefont{Shender}(1982)}]{shend82}
\bibinfo{author}{\bibfnamefont{E.~F.} \bibnamefont{Shender}},
  \bibinfo{journal}{Sov. Phys. JETP} \textbf{\bibinfo{volume}{56}},
  \bibinfo{pages}{178} (\bibinfo{year}{1982}).


\bibitem[{\citenamefont{Melzi}(2000)}]{melzi00}
\bibinfo{author}{\bibfnamefont{R.}~\bibnamefont{Melzi}},
  \bibinfo{author}{\bibfnamefont{P.}~\bibnamefont{Carretta}},
  \bibinfo{author}{\bibfnamefont{A.}~\bibnamefont{Lascialfari}},
\bibinfo{author}{\bibfnamefont{M.}~\bibnamefont{Mambrini}},
  \bibinfo{author}{\bibfnamefont{M.}~\bibnamefont{Troyer}},
  \bibinfo{author}{\bibfnamefont{P.}~\bibnamefont{Millet}}, \bibnamefont{and}
  \bibinfo{author}{\bibfnamefont{F.}~\bibnamefont{Mila}},
  \bibinfo{journal}{Phys. Rev. Lett.} \textbf{\bibinfo{volume}{85}},
  \bibinfo{pages}{1318} (\bibinfo{year}{2000}).


\bibitem[{\citenamefont{Rosner}(2002)}]{rosner02}
\bibinfo{author}{\bibfnamefont{H.}~\bibnamefont{Rosner}},
  \bibinfo{author}{\bibfnamefont{R.~R.~P.}~\bibnamefont{Singh}},
  \bibinfo{author}{\bibfnamefont{W.~H.}~\bibnamefont{Zheng}},
\bibinfo{author}{\bibfnamefont{J.}~\bibnamefont{Oitmaa}},
  \bibinfo{author}{\bibfnamefont{S.~L.}~\bibnamefont{Drechsler}}, \bibnamefont{and}
  \bibinfo{author}{\bibfnamefont{W.~E.}~\bibnamefont{Pickett}},
  \bibinfo{journal}{Phys. Rev. Lett.} \textbf{\bibinfo{volume}{88}},
  \bibinfo{pages}{186405} (\bibinfo{year}{2002}).


\bibitem[{\citenamefont{Schmalfu\ss{}}(2006)}]{schma06}
\bibinfo{author}{\bibfnamefont{D.} \bibnamefont{Schmalfu\ss{}}},
\bibinfo{author}{\bibfnamefont{R.} \bibnamefont{Darradi}},
\bibinfo{author}{\bibfnamefont{J.} \bibnamefont{Richter}}, 
\bibinfo{author}{\bibfnamefont{J.} \bibnamefont{Schulenburg}}, \bibnamefont{and}
  \bibinfo{author}{\bibfnamefont{D.}~\bibnamefont{Ihle}},
  \bibinfo{journal}{Phys. Rev. Lett.} \textbf{\bibinfo{volume}{97}},
  \bibinfo{pages}{157201} (\bibinfo{year}{2006}).


\bibitem[{\citenamefont{Nunes}(2010)}]{nunes10}
\bibinfo{author}{\bibfnamefont{W.~A.} \bibnamefont{Nunes}},
\bibinfo{author}{\bibfnamefont{J.~R.} \bibnamefont{de Sousa}},
\bibinfo{author}{\bibfnamefont{J.~R.} \bibnamefont{Viana}}, \bibnamefont{and}
  \bibinfo{author}{\bibfnamefont{J.}~\bibnamefont{Richter}},
  \bibinfo{journal}{J. Phys. Condens. Matter} \textbf{\bibinfo{volume}{22}},
  \bibinfo{pages}{146004} (\bibinfo{year}{2010}).


\bibitem[{\citenamefont{Nunes}(2010)}]{nunes10a}
\bibinfo{author}{\bibfnamefont{W.~A.} \bibnamefont{Nunes}},
\bibinfo{author}{\bibfnamefont{J.~R.} \bibnamefont{Viana}}, \bibnamefont{and}
  \bibinfo{author}{\bibfnamefont{J.~R.} \bibnamefont{de Sousa}},
 p. \bibinfo{pages}{in press} (\bibinfo{year}{2010}).


\bibitem[{\citenamefont{Majumdar}(2010)}]{maj10}
  \bibinfo{author}{\bibfnamefont{K.}~\bibnamefont{Majumdar}},
  \bibinfo{journal}{J. Phys. Condens. Matter} \textbf{\bibinfo{volume}{23}},
  \bibinfo{pages}{46001} (\bibinfo{year}{2011}).


\bibitem[{\citenamefont{Kamihara et~al.}(2008)\citenamefont{Kamihara, Watanabe,Hirano, and Hosono}}]{kamih08}
\bibinfo{author}{\bibfnamefont{Y.}~\bibnamefont{Kamihara}},
  \bibinfo{author}{\bibfnamefont{T.}~\bibnamefont{Watanabe}},
  \bibinfo{author}{\bibfnamefont{M.}~\bibnamefont{Hirano}}, \bibnamefont{and}
  \bibinfo{author}{\bibfnamefont{H.}~\bibnamefont{Hosono}},
  \bibinfo{journal}{J. Am. Chem. Soc.} \textbf{\bibinfo{volume}{130}},
  \bibinfo{pages}{3296} (\bibinfo{year}{2008}).


\bibitem[{\citenamefont{Fang et~al.}(2008)\citenamefont{Fang, Yao, Tsai, Hu, and Kivelson}}]{fang08}
\bibinfo{author}{\bibfnamefont{C.}~\bibnamefont{Fang}},
  \bibinfo{author}{\bibfnamefont{H.}~\bibnamefont{Yao}},
  \bibinfo{author}{\bibfnamefont{W.~F.}~\bibnamefont{Tsai}},
  \bibinfo{author}{\bibfnamefont{J.~P.}~\bibnamefont{Hu}}, \bibnamefont{and}
  \bibinfo{author}{\bibfnamefont{S.~A.} \bibnamefont{Kivelson}},
  \bibinfo{journal}{Phys. Rev. B} \textbf{\bibinfo{volume}{77}},
  \bibinfo{pages}{224509} (\bibinfo{year}{2008}).


\bibitem[{\citenamefont{Xu et~al.}(2008)\citenamefont{Xu, M\"uller, and Sachdev}}]{xu08}
\bibinfo{author}{\bibfnamefont{C.}~\bibnamefont{Xu}},
  \bibinfo{author}{\bibfnamefont{M.}~\bibnamefont{M\"uller}}, \bibnamefont{and}
  \bibinfo{author}{\bibfnamefont{S.}~\bibnamefont{Sachdev}},
  \bibinfo{journal}{Phys. Rev. B} \textbf{\bibinfo{volume}{78}},
  \bibinfo{pages}{020501(R)} (\bibinfo{year}{2008}).


\bibitem[{\citenamefont{Ma et~al.}(2009)\citenamefont{Ma, Ji, Hu, Lu, and Xiang}}]{ma09}
\bibinfo{author}{\bibfnamefont{F.}~\bibnamefont{Ma}},
  \bibinfo{author}{\bibfnamefont{W.}~\bibnamefont{Ji}},
  \bibinfo{author}{\bibfnamefont{J.}~\bibnamefont{Hu}},
  \bibinfo{author}{\bibfnamefont{Z.-Y.} \bibnamefont{Lu}}, \bibnamefont{and}
  \bibinfo{author}{\bibfnamefont{T.}~\bibnamefont{Xiang}},
  \bibinfo{journal}{Phys. Rev. Lett.} \textbf{\bibinfo{volume}{102}},
  \bibinfo{pages}{177003} (\bibinfo{year}{2009}).


\bibitem[{\citenamefont{Uhrig et~al.}(2009)\citenamefont{Uhrig, Holt, Oitmaa, Sushkov, and Singh}}]{uhrig09a}
\bibinfo{author}{\bibfnamefont{G.~S.} \bibnamefont{Uhrig}},
  \bibinfo{author}{\bibfnamefont{M.}~\bibnamefont{Holt}},
  \bibinfo{author}{\bibfnamefont{J.}~\bibnamefont{Oitmaa}},
  \bibinfo{author}{\bibfnamefont{O.~P.} \bibnamefont{Sushkov}},
  \bibnamefont{and} \bibinfo{author}{\bibfnamefont{R.~R.~P.}~\bibnamefont{Singh}},
  \bibinfo{journal}{Phys. Rev. B} \textbf{\bibinfo{volume}{79}},
  \bibinfo{pages}{092416} (\bibinfo{year}{2009}).


\bibitem[{\citenamefont{Bao et~al.}(2009)\citenamefont{Bao, Qiu, Huang, Green, Zajdel, Fitzsimmons, Zhernenkov, Chang, Fang, Qian et~al.}}]{bao09}
\bibinfo{author}{\bibfnamefont{W.}~\bibnamefont{Bao}},
  \bibinfo{author}{\bibfnamefont{Y.}~\bibnamefont{Qiu}},
  \bibinfo{author}{\bibfnamefont{Q.}~\bibnamefont{Huang}},
  \bibinfo{author}{\bibfnamefont{M.~A.} \bibnamefont{Green}},
  \bibinfo{author}{\bibfnamefont{P.}~\bibnamefont{Zajdel}},
  \bibinfo{author}{\bibfnamefont{M.~R.} \bibnamefont{Fitzsimmons}},
  \bibinfo{author}{\bibfnamefont{M.}~\bibnamefont{Zhernenkov}},
  \bibinfo{author}{\bibfnamefont{S.}~\bibnamefont{Chang}},
  \bibinfo{author}{\bibfnamefont{M.}~\bibnamefont{Fang}},
  \bibinfo{author}{\bibfnamefont{B.}~\bibnamefont{Qian}}, \bibnamefont{et~al.},
  \bibinfo{journal}{Phys. Rev. Lett.} \textbf{\bibinfo{volume}{102}},
  \bibinfo{pages}{247001} (\bibinfo{year}{2009}).


\bibitem[{\citenamefont{Pulikkotil et~al.}(2010)\citenamefont{Pulikkotil, Ke, van Schilfgaarde, Kotani, and V.P.Antropov}}]{pulik10}
\bibinfo{author}{\bibfnamefont{J.}~\bibnamefont{Pulikkotil}},
  \bibinfo{author}{\bibfnamefont{L.}~\bibnamefont{Ke}},
  \bibinfo{author}{\bibfnamefont{M.}~\bibnamefont{van Schilfgaarde}},
  \bibinfo{author}{\bibfnamefont{T.}~\bibnamefont{Kotani}}, \bibnamefont{and}
  \bibinfo{author}{\bibnamefont{V.P.Antropov}}, p. \bibinfo{pages}{arXiv:0809.0283v2}
  (\bibinfo{year}{2010}).


\bibitem[{\citenamefont{Si and Abrahams}(2008)}]{si08}
\bibinfo{author}{\bibfnamefont{Q.}~\bibnamefont{Si}} \bibnamefont{and}
  \bibinfo{author}{\bibfnamefont{E.}~\bibnamefont{Abrahams}},
  \bibinfo{journal}{Phys. Rev. Lett.} \textbf{\bibinfo{volume}{101}},
  \bibinfo{pages}{076401} (\bibinfo{year}{2008}).


\bibitem[{\citenamefont{Wu et~al.}(2008)\citenamefont{Wu, Phillips, and CastroNeto}}]{wu08}
\bibinfo{author}{\bibfnamefont{J.}~\bibnamefont{Wu}},
  \bibinfo{author}{\bibfnamefont{P.}~\bibnamefont{Phillips}}, \bibnamefont{and}
  \bibinfo{author}{\bibfnamefont{A.~H.} \bibnamefont{Castro Neto}},
  \bibinfo{journal}{Phys. Rev. Lett.} \textbf{\bibinfo{volume}{101}},
  \bibinfo{pages}{126401} (\bibinfo{year}{2008}).


\bibitem[{\citenamefont{Zhao et~al.}(2008)\citenamefont{Zhao, Yao, Li, Hong, Chen, Chang, II, Lynn, Mook, Chen et~al.}}]{zhao08a}
\bibinfo{author}{\bibfnamefont{J.}~\bibnamefont{Zhao}},
  \bibinfo{author}{\bibfnamefont{D.-X.} \bibnamefont{Yao}},
  \bibinfo{author}{\bibfnamefont{S.}~\bibnamefont{Li}},
  \bibinfo{author}{\bibfnamefont{T.}~\bibnamefont{Hong}},
  \bibinfo{author}{\bibfnamefont{Y.}~\bibnamefont{Chen}},
  \bibinfo{author}{\bibfnamefont{S.}~\bibnamefont{Chang}},
  \bibinfo{author}{\bibfnamefont{W.~R.} \bibnamefont{II}},
  \bibinfo{author}{\bibfnamefont{J.~W.} \bibnamefont{Lynn}},
  \bibinfo{author}{\bibfnamefont{H.~A.} \bibnamefont{Mook}},
  \bibinfo{author}{\bibfnamefont{G.~F.} \bibnamefont{Chen}},
  \bibnamefont{et~al.}, \bibinfo{journal}{Phys. Rev. Lett.}
  \textbf{\bibinfo{volume}{101}}, \bibinfo{pages}{167203}
  (\bibinfo{year}{2008}).


\bibitem[{\citenamefont{Zhao et~al.}(2009)\citenamefont{Zhao, Adroja, Yao, Bewley, Li, Wang, Wu, Chen, Hu, and Dai}}]{zhao09}
\bibinfo{author}{\bibfnamefont{J.}~\bibnamefont{Zhao}},
  \bibinfo{author}{\bibfnamefont{D.~T.} \bibnamefont{Adroja}},
  \bibinfo{author}{\bibfnamefont{D.-X.} \bibnamefont{Yao}},
  \bibinfo{author}{\bibfnamefont{R.}~\bibnamefont{Bewley}},
  \bibinfo{author}{\bibfnamefont{S.}~\bibnamefont{Li}},
  \bibinfo{author}{\bibfnamefont{X.~F.} \bibnamefont{Wang}},
  \bibinfo{author}{\bibfnamefont{G.}~\bibnamefont{Wu}},
  \bibinfo{author}{\bibfnamefont{X.~H.} \bibnamefont{Chen}},
  \bibinfo{author}{\bibfnamefont{J.}~\bibnamefont{Hu}}, \bibnamefont{and}
  \bibinfo{author}{\bibfnamefont{P.}~\bibnamefont{Dai}},
  \bibinfo{journal}{Nature Phys.} \textbf{\bibinfo{volume}{5}},
  \bibinfo{pages}{555} (\bibinfo{year}{2009}).


\bibitem[{\citenamefont{Ewings et~al.}(2008)\citenamefont{Ewings, Perring, Bewley, Guidi, Pitcher, Parker, Clarke, and Boothroyd}}]{ewing08}
\bibinfo{author}{\bibfnamefont{R.~A.} \bibnamefont{Ewings}},
  \bibinfo{author}{\bibfnamefont{T.~G.} \bibnamefont{Perring}},
  \bibinfo{author}{\bibfnamefont{R.~I.} \bibnamefont{Bewley}},
  \bibinfo{author}{\bibfnamefont{T.}~\bibnamefont{Guidi}},
  \bibinfo{author}{\bibfnamefont{M.~J.} \bibnamefont{Pitcher}},
  \bibinfo{author}{\bibfnamefont{D.~R.} \bibnamefont{Parker}},
  \bibinfo{author}{\bibfnamefont{S.~J.} \bibnamefont{Clarke}},
  \bibnamefont{and} \bibinfo{author}{\bibfnamefont{A.~T.}
  \bibnamefont{Boothroyd}}, \bibinfo{journal}{Phys. Rev. B}
  \textbf{\bibinfo{volume}{78}}, \bibinfo{pages}{220501(R)}
  (\bibinfo{year}{2008}).


\bibitem[{\citenamefont{McQueeney et~al.}(2008)\citenamefont{McQueeney, Diallo, Antropov, Samolyuk, Broholm, Ni, Nandi, Yethiraj, Zarestky, Pulikkotil et~al.}}]{mcque08}
\bibinfo{author}{\bibfnamefont{R.~J.} \bibnamefont{McQueeney}},
  \bibinfo{author}{\bibfnamefont{S.~O.} \bibnamefont{Diallo}},
  \bibinfo{author}{\bibfnamefont{V.~P.} \bibnamefont{Antropov}},
  \bibinfo{author}{\bibfnamefont{G.}~\bibnamefont{Samolyuk}},
  \bibinfo{author}{\bibfnamefont{C.}~\bibnamefont{Broholm}},
  \bibinfo{author}{\bibfnamefont{N.}~\bibnamefont{Ni}},
  \bibinfo{author}{\bibfnamefont{S.}~\bibnamefont{Nandi}},
  \bibinfo{author}{\bibfnamefont{M.}~\bibnamefont{Yethiraj}},
  \bibinfo{author}{\bibfnamefont{J.~L.} \bibnamefont{Zarestky}},
  \bibinfo{author}{\bibfnamefont{J.~J.} \bibnamefont{Pulikkotil}},
  \bibnamefont{et~al.}, \bibinfo{journal}{Phys. Rev. Lett.}
  \textbf{\bibinfo{volume}{101}}, \bibinfo{pages}{227205}
  (\bibinfo{year}{2008}).


\bibitem[{\citenamefont{de~la Cruz et~al.}(2008)\citenamefont{de~la Cruz, Huang, Lynn, Li, II, Zarestky, Mook, Chen, Luo, Wang et~al.}}]{cruz08}
\bibinfo{author}{\bibfnamefont{C.}~\bibnamefont{de~la Cruz}},
  \bibinfo{author}{\bibfnamefont{Q.}~\bibnamefont{Huang}},
  \bibinfo{author}{\bibfnamefont{J.~W.} \bibnamefont{Lynn}},
  \bibinfo{author}{\bibfnamefont{J.}~\bibnamefont{Li}},
  \bibinfo{author}{\bibfnamefont{W.~R.} \bibnamefont{II}},
  \bibinfo{author}{\bibfnamefont{J.~L.} \bibnamefont{Zarestky}},
  \bibinfo{author}{\bibfnamefont{H.~A.} \bibnamefont{Mook}},
  \bibinfo{author}{\bibfnamefont{G.~F.} \bibnamefont{Chen}},
  \bibinfo{author}{\bibfnamefont{J.~L.} \bibnamefont{Luo}},
  \bibinfo{author}{\bibfnamefont{N.~L.} \bibnamefont{Wang}},
  \bibnamefont{et~al.}, \bibinfo{journal}{Nature}
  \textbf{\bibinfo{volume}{453}}, \bibinfo{pages}{899} (\bibinfo{year}{2008}).


\bibitem[{\citenamefont{Drew et~al.}(2008)\citenamefont{Drew, Pratt, Lancaster, Blundell, Baker, Liu, Wu, Chen, Watanabe, Malik et~al.}}]{drew08}
\bibinfo{author}{\bibfnamefont{A.~J.} \bibnamefont{Drew}},
  \bibinfo{author}{\bibfnamefont{F.~L.} \bibnamefont{Pratt}},
  \bibinfo{author}{\bibfnamefont{T.}~\bibnamefont{Lancaster}},
  \bibinfo{author}{\bibfnamefont{S.~J.} \bibnamefont{Blundell}},
  \bibinfo{author}{\bibfnamefont{P.~J.} \bibnamefont{Baker}},
  \bibinfo{author}{\bibfnamefont{R.~H.} \bibnamefont{Liu}},
  \bibinfo{author}{\bibfnamefont{G.}~\bibnamefont{Wu}},
  \bibinfo{author}{\bibfnamefont{X.~H.} \bibnamefont{Chen}},
  \bibinfo{author}{\bibfnamefont{I.}~\bibnamefont{Watanabe}},
  \bibinfo{author}{\bibfnamefont{V.~K.} \bibnamefont{Malik}},
  \bibnamefont{et~al.}, \bibinfo{journal}{Phys. Rev. Lett.}
  \textbf{\bibinfo{volume}{101}}, \bibinfo{pages}{097010}
  (\bibinfo{year}{2008}).


\bibitem[{\citenamefont{Klauss et~al.}(2008)\citenamefont{Klauss, Luekens, Klingeler, Hess, Litterst, Kraken, Korshunov, Eremin, Drechsler, Khasanov et~al.}}]{klaus08}
\bibinfo{author}{\bibfnamefont{H.-H.} \bibnamefont{Klauss}},
  \bibinfo{author}{\bibfnamefont{H.}~\bibnamefont{Luekens}},
  \bibinfo{author}{\bibfnamefont{R.}~\bibnamefont{Klingeler}},
  \bibinfo{author}{\bibfnamefont{C.}~\bibnamefont{Hess}},
  \bibinfo{author}{\bibfnamefont{F.~J.} \bibnamefont{Litterst}},
  \bibinfo{author}{\bibfnamefont{M.}~\bibnamefont{Kraken}},
  \bibinfo{author}{\bibfnamefont{M.~M.} \bibnamefont{Korshunov}},
  \bibinfo{author}{\bibfnamefont{I.}~\bibnamefont{Eremin}},
  \bibinfo{author}{\bibfnamefont{S.-L.} \bibnamefont{Drechsler}},
  \bibinfo{author}{\bibfnamefont{R.}~\bibnamefont{Khasanov}},
  \bibnamefont{et~al.}, \bibinfo{journal}{Phys. Rev. Lett.}
  \textbf{\bibinfo{volume}{101}}, \bibinfo{pages}{077005}
  (\bibinfo{year}{2008}).


\bibitem[{\citenamefont{Luetkens et~al.}(2009)\citenamefont{Luetkens, Klauss, Kraken, Litterst, Dellmann, Klingeler, Hess, Khasanov, Amato, Baines et~al.}}]{luetk09}
\bibinfo{author}{\bibfnamefont{H.}~\bibnamefont{Luetkens}},
  \bibinfo{author}{\bibfnamefont{H.-H.} \bibnamefont{Klauss}},
  \bibinfo{author}{\bibfnamefont{M.}~\bibnamefont{Kraken}},
  \bibinfo{author}{\bibfnamefont{F.~J.} \bibnamefont{Litterst}},
  \bibinfo{author}{\bibfnamefont{T.}~\bibnamefont{Dellmann}},
  \bibinfo{author}{\bibfnamefont{R.}~\bibnamefont{Klingeler}},
  \bibinfo{author}{\bibfnamefont{C.}~\bibnamefont{Hess}},
  \bibinfo{author}{\bibfnamefont{R.}~\bibnamefont{Khasanov}},
  \bibinfo{author}{\bibfnamefont{A.}~\bibnamefont{Amato}},
  \bibinfo{author}{\bibfnamefont{C.}~\bibnamefont{Baines}},
  \bibnamefont{et~al.}, \bibinfo{journal}{Nature Mat.}
  \textbf{\bibinfo{volume}{2397}} (\bibinfo{year}{2009}).


\bibitem[{\citenamefont{Huang et~al.}(2008)\citenamefont{Huang, Qiu, Bao, Green, Lynn, Gasparovic, Wu, Wu, and Chen}}]{huang08}
\bibinfo{author}{\bibfnamefont{Q.}~\bibnamefont{Huang}},
  \bibinfo{author}{\bibfnamefont{Y.}~\bibnamefont{Qiu}},
  \bibinfo{author}{\bibfnamefont{W.}~\bibnamefont{Bao}},
  \bibinfo{author}{\bibfnamefont{M.~A.} \bibnamefont{Green}},
  \bibinfo{author}{\bibfnamefont{J.~W.}~\bibnamefont{Lynn}},
  \bibinfo{author}{\bibfnamefont{Y.~C.} \bibnamefont{Gasparovic}},
  \bibinfo{author}{\bibfnamefont{T.}~\bibnamefont{Wu}},
  \bibinfo{author}{\bibfnamefont{G.}~\bibnamefont{Wu}}, \bibnamefont{and}
  \bibinfo{author}{\bibfnamefont{X.~H.} \bibnamefont{Chen}},
  \bibinfo{journal}{Phys. Rev. Lett.} \textbf{\bibinfo{volume}{101}},
  \bibinfo{pages}{257003} (\bibinfo{year}{2008}).


\bibitem[{\citenamefont{Zhao et~al.}(2008)}]{zhao08b}
\bibinfo{author}{\bibfnamefont{J.}~\bibnamefont{Zhao}},
\bibinfo{author}{\bibfnamefont{Q.}~\bibnamefont{Huang}},
\bibinfo{author}{\bibfnamefont{C.}~\bibnamefont{de la Cruz}},
\bibinfo{author}{\bibfnamefont{S.} \bibnamefont{Li}}, 
\bibinfo{author}{\bibfnamefont{J.~W.}~\bibnamefont{Lynn}},
\bibinfo{author}{\bibfnamefont{Y.} \bibnamefont{Chen}},
\bibinfo{author}{\bibfnamefont{M.~A.}~\bibnamefont{Green}},
\bibinfo{author}{\bibfnamefont{G.~F.}~\bibnamefont{Chen}}
\bibinfo{author}{\bibfnamefont{G.} \bibnamefont{Li}},
\bibinfo{author}{\bibfnamefont{Z.}~\bibnamefont{Li}},
\bibinfo{author}{\bibfnamefont{J.~L.}~\bibnamefont{Luo}}
\bibinfo{author}{\bibfnamefont{N.~L.} \bibnamefont{Wang}}, \bibnamefont{and}
\bibinfo{author}{\bibfnamefont{P.} \bibnamefont{Dai}},
\bibinfo{journal}{Nature. Mat.} \textbf{\bibinfo{volume}{7}},
\bibinfo{pages}{953} (\bibinfo{year}{2008}).


\bibitem[{\citenamefont{Ong et~al.}(2009)\citenamefont{Ong, Uhrig, and Sushkov}}]{ong09}
\bibinfo{author}{\bibfnamefont{A.}~\bibnamefont{Ong}},
  \bibinfo{author}{\bibfnamefont{G.~S.} \bibnamefont{Uhrig}}, \bibnamefont{and}
  \bibinfo{author}{\bibfnamefont{O.~P.} \bibnamefont{Sushkov}},
  \bibinfo{journal}{Phys. Rev. B} \textbf{\bibinfo{volume}{80}},
  \bibinfo{pages}{014514} (\bibinfo{year}{2009}).


\bibitem[{\citenamefont{Dong et~al.}(2008)\citenamefont{Dong, Zhand, Xu, Li,  Li, Hu, Wu, Chen, Dai, Luo et~al.}}]{dong08b}
\bibinfo{author}{\bibfnamefont{J.}~\bibnamefont{Dong}},
  \bibinfo{author}{\bibfnamefont{H.~J.} \bibnamefont{Zhand}},
  \bibinfo{author}{\bibfnamefont{G.}~\bibnamefont{Xu}},
  \bibinfo{author}{\bibfnamefont{Z.}~\bibnamefont{Li}},
  \bibinfo{author}{\bibfnamefont{G.}~\bibnamefont{Li}},
  \bibinfo{author}{\bibfnamefont{W.~Z.} \bibnamefont{Hu}},
  \bibinfo{author}{\bibfnamefont{D.}~\bibnamefont{Wu}},
  \bibinfo{author}{\bibfnamefont{G.~F.} \bibnamefont{Chen}},
  \bibinfo{author}{\bibfnamefont{X.}~\bibnamefont{Dai}},
  \bibinfo{author}{\bibfnamefont{J.~L.} \bibnamefont{Luo}},
  \bibnamefont{et~al.}, \bibinfo{journal}{Europhys. Lett.}
  \textbf{\bibinfo{volume}{83}}, \bibinfo{pages}{27006} (\bibinfo{year}{2008}).


\bibitem[{\citenamefont{Z.P.Yin et~al.}(2008)\citenamefont{Z.P.Yin, Leb\`egue, Han, Neal, Savrasov, and Pickett}}]{yin08}
\bibinfo{author}{\bibnamefont{Z.P.Yin}},
  \bibinfo{author}{\bibfnamefont{S.}~\bibnamefont{Leb\`egue}},
  \bibinfo{author}{\bibfnamefont{M.~J.} \bibnamefont{Han}},
  \bibinfo{author}{\bibfnamefont{B.~P.} \bibnamefont{Neal}},
  \bibinfo{author}{\bibfnamefont{S.~Y.} \bibnamefont{Savrasov}},
  \bibnamefont{and} \bibinfo{author}{\bibfnamefont{W.~E.}
  \bibnamefont{Pickett}}, \bibinfo{journal}{Phys. Rev. Lett.}
  \textbf{\bibinfo{volume}{101}}, \bibinfo{pages}{047001}
  (\bibinfo{year}{2008}).


\bibitem[{\citenamefont{Cao et~al.}(2008)\citenamefont{Cao, Hirschfeld, and Cheng}}]{cao08}
\bibinfo{author}{\bibfnamefont{C.}~\bibnamefont{Cao}},
  \bibinfo{author}{\bibfnamefont{P.~J.} \bibnamefont{Hirschfeld}},
  \bibnamefont{and} \bibinfo{author}{\bibfnamefont{H.-P.} \bibnamefont{Cheng}},
  \bibinfo{journal}{Phys. Rev. B} \textbf{\bibinfo{volume}{77}},
  \bibinfo{pages}{220506} (\bibinfo{year}{2008}).


\bibitem[{\citenamefont{Yao and Carlson}(2008)}]{yao08}
\bibinfo{author}{\bibfnamefont{D.-X.} \bibnamefont{Yao}} \bibnamefont{and}
  \bibinfo{author}{\bibfnamefont{E.~W.} \bibnamefont{Carlson}},
  \bibinfo{journal}{Phys. Rev. B} \textbf{\bibinfo{volume}{78}},
  \bibinfo{pages}{052507} (\bibinfo{year}{2008}).


\bibitem[{\citenamefont{Yildrim}(2008)}]{yildr08}
\bibinfo{author}{\bibfnamefont{T.}~\bibnamefont{Yildirim}},
  \bibinfo{journal}{Phys. Rev. Lett.} \textbf{\bibinfo{volume}{101}},
  \bibinfo{pages}{057010} (\bibinfo{year}{2008}).


\bibitem[{\citenamefont{Smerald and Shannon}(2009)}]{smera10}
\bibinfo{author}{\bibfnamefont{A.}~\bibnamefont{Smerald}} \bibnamefont{and}
  \bibinfo{author}{\bibfnamefont{N.}~\bibnamefont{Shannon}}, 
   \bibinfo{journal}{Europhys. Lett.} \textbf{\bibinfo{volume}{92}},
  \bibinfo{pages}{47005} (\bibinfo{year}{2010}).


\bibitem[{\citenamefont{Stanek et~al.}(2010)\citenamefont{Stanek, Holt, Sushkov, and Uhrig}}]{stane10}
\bibinfo{author}{\bibfnamefont{D.}~\bibnamefont{Stanek}},
  \bibinfo{author}{\bibfnamefont{M.}~\bibnamefont{Holt}},
  \bibinfo{author}{\bibfnamefont{O.~P.} \bibnamefont{Sushkov}},
  \bibnamefont{and} \bibinfo{author}{\bibfnamefont{G.~S.} \bibnamefont{Uhrig}},
  p. \bibinfo{pages}{in preparation} (\bibinfo{year}{2011}).


\bibitem[{\citenamefont{Oitmaa and Weihong}(1996)}]{oitma96c}
\bibinfo{author}{\bibfnamefont{J.}~\bibnamefont{Oitmaa}} \bibnamefont{and}
  \bibinfo{author}{\bibfnamefont{W.~H.}~\bibnamefont{Zheng}},
  \bibinfo{journal}{Phys. Rev. B} \textbf{\bibinfo{volume}{54}},
  \bibinfo{pages}{3022} (\bibinfo{year}{1996}).


\bibitem[{\citenamefont{Dyson}(1956)}]{dyson56a}
\bibinfo{author}{\bibfnamefont{F.~J.} \bibnamefont{Dyson}},
  \bibinfo{journal}{Phys. Rev.} \textbf{\bibinfo{volume}{102}},
  \bibinfo{pages}{1217} (\bibinfo{year}{1956}).


\bibitem[{\citenamefont{Maleev}(1957)}]{malee57}
\bibinfo{author}{\bibfnamefont{S.~V.} \bibnamefont{Maleev}},
  \bibinfo{journal}{Zh. Eksp. Teor. Fiz.} \textbf{\bibinfo{volume}{33}},
  \bibinfo{pages}{1010} (\bibinfo{year}{1957}).


\bibitem[{\citenamefont{Auerbach}(1994)}]{auerb94}
\bibinfo{author}{\bibfnamefont{A.}~\bibnamefont{Auerbach}},
  \emph{\bibinfo{title}{Interacting Electrons and Quantum Magnetism}}, Graduate
  Texts in Contemporary Physics (\bibinfo{publisher}{Springer},
  \bibinfo{address}{New York}, \bibinfo{year}{1994}).


\bibitem[{\citenamefont{Bishop et~al.}(2008)\citenamefont{Bishop, Li, Darradi, and Richter}}]{bishop08}
\bibinfo{author}{\bibfnamefont{R.~F.} \bibnamefont{Bishop}},
  \bibinfo{author}{\bibfnamefont{P.~H.~Y.}~\bibnamefont{Li}},
  \bibinfo{author}{\bibfnamefont{R.} \bibnamefont{Darradi}},
  \bibnamefont{and} \bibinfo{author}{\bibfnamefont{J.}
  \bibnamefont{Richter}}, \bibinfo{journal}{Europhys. Lett.}
  \textbf{\bibinfo{volume}{83}}, \bibinfo{pages}{47004}
  (\bibinfo{year}{2008}).


\bibitem[{\citenamefont{Han et~al.}(2009)\citenamefont{Han, Yin, Pickett, and Savrasov}}]{han09}
\bibinfo{author}{\bibfnamefont{M.~J.} \bibnamefont{Han}},
  \bibinfo{author}{\bibfnamefont{Q.}~\bibnamefont{Yin}},
  \bibinfo{author}{\bibfnamefont{W.~E.} \bibnamefont{Pickett}},
  \bibnamefont{and} \bibinfo{author}{\bibfnamefont{S.~Y.}
  \bibnamefont{Savrasov}}, \bibinfo{journal}{Phys. Rev. Lett.}
  \textbf{\bibinfo{volume}{102}}, \bibinfo{pages}{107003}
  (\bibinfo{year}{2009}).


\bibitem[{\citenamefont{Chen et~al.}(2008)\citenamefont{Chen, Lynn, Li, Li, Chen, Luo, Wang, Dai, dela Cruz, and Mook}}]{chen08}
\bibinfo{author}{\bibfnamefont{Y.}~\bibnamefont{Chen}},
  \bibinfo{author}{\bibfnamefont{J.~W.} \bibnamefont{Lynn}},
  \bibinfo{author}{\bibfnamefont{J.}~\bibnamefont{Li}},
  \bibinfo{author}{\bibfnamefont{G.}~\bibnamefont{Li}},
  \bibinfo{author}{\bibfnamefont{G.~F.} \bibnamefont{Chen}},
  \bibinfo{author}{\bibfnamefont{J.~L.} \bibnamefont{Luo}},
  \bibinfo{author}{\bibfnamefont{N.~L.} \bibnamefont{Wang}},
  \bibinfo{author}{\bibfnamefont{P.}~\bibnamefont{Dai}},
  \bibinfo{author}{\bibfnamefont{C.}~\bibnamefont{dela Cruz}},
  \bibnamefont{and} \bibinfo{author}{\bibfnamefont{H.~A.} \bibnamefont{Mook}},
  \bibinfo{journal}{Phys. Rev. B} \textbf{\bibinfo{volume}{78}},
  \bibinfo{pages}{064515} (\bibinfo{year}{2008}).

\bibitem[{\citenamefont{Diallo et~al.}(2009)\citenamefont{Diallo, Antropov, Perring, Broholm, Pulikkotil, Ni, Bud'ko, Canfield, Kreyssig, Goldman et~al.}}]{diall09}
\bibinfo{author}{\bibfnamefont{S.~O.} \bibnamefont{Diallo}},
  \bibinfo{author}{\bibfnamefont{V.~P.} \bibnamefont{Antropov}},
  \bibinfo{author}{\bibfnamefont{T.~G.} \bibnamefont{Perring}},
  \bibinfo{author}{\bibfnamefont{C.}~\bibnamefont{Broholm}},
  \bibinfo{author}{\bibfnamefont{J.~J.} \bibnamefont{Pulikkotil}},
  \bibinfo{author}{\bibfnamefont{N.}~\bibnamefont{Ni}},
  \bibinfo{author}{\bibfnamefont{S.~L.} \bibnamefont{Bud'ko}},
  \bibinfo{author}{\bibfnamefont{P.~C.} \bibnamefont{Canfield}},
  \bibinfo{author}{\bibfnamefont{A.}~\bibnamefont{Kreyssig}},
  \bibinfo{author}{\bibfnamefont{A.~I.} \bibnamefont{Goldman}},
  \bibnamefont{et~al.}, \bibinfo{journal}{Phys. Rev. Lett.}
  \textbf{\bibinfo{volume}{102}}, \bibinfo{pages}{187206}(\bibinfo{year}{2009}).


\bibitem[{\citenamefont{Kaneko et~al.}(2008)\citenamefont{Kaneko, Hoser, Caroca-Canales, Jesche, Krellner, Stockert, and Geibel}}]{kanek08}
\bibinfo{author}{\bibfnamefont{K.}~\bibnamefont{Kaneko}},
  \bibinfo{author}{\bibfnamefont{A.}~\bibnamefont{Hoser}},
  \bibinfo{author}{\bibfnamefont{N.}~\bibnamefont{Caroca-Canales}},
  \bibinfo{author}{\bibfnamefont{A.}~\bibnamefont{Jesche}},
  \bibinfo{author}{\bibfnamefont{C.}~\bibnamefont{Krellner}},
  \bibinfo{author}{\bibfnamefont{O.}~\bibnamefont{Stockert}}, \bibnamefont{and}
  \bibinfo{author}{\bibfnamefont{C.}~\bibnamefont{Geibel}},
  \bibinfo{journal}{Phys. Rev. B} \textbf{\bibinfo{volume}{78}},
  \bibinfo{pages}{212502} (\bibinfo{year}{2008}).


\bibitem[{\citenamefont{Lv et~al.}(2010)\citenamefont{Lv, Kr\"uger, and Phillips}}]{lv10}
\bibinfo{author}{\bibfnamefont{W.}~\bibnamefont{Lv}},
  \bibinfo{author}{\bibfnamefont{F.}~\bibnamefont{Kr\"uger}}, \bibnamefont{and}
  \bibinfo{author}{\bibfnamefont{P.}~\bibnamefont{Phillips}},
  \bibinfo{journal}{Phys. Rev. B} \textbf{\bibinfo{volume}{82}},
  \bibinfo{pages}{045125} (\bibinfo{year}{2010}).


\end{thebibliography}
\end{document}